\documentclass[aps,
pre,
superscriptaddress,
amsmath,amssymb,
reprint,
showkeys,
floatfix]{revtex4-2}

\usepackage{algorithm}
\usepackage{algpseudocode}
\usepackage{array}
\usepackage{mathtools}
\usepackage[dvipsnames]{xcolor}
\usepackage{lipsum}
\usepackage{tikz}
\usepackage{wrapfig}
\usepackage[T1]{fontenc}
\usepackage[utf8]{inputenc}
\usepackage{amsmath}
\usepackage{amssymb}
\usepackage{graphicx}
\usepackage{refstyle}
\usepackage{multirow}
\usepackage{blkarray,booktabs,bigstrut}
\usepackage{placeins}
\usepackage{makecell}
\usepackage{soul}
\usepackage[
citecolor=blue,
colorlinks,
linkcolor=blue,
urlcolor=blue,
]{hyperref}

\usepackage{xcolor}
\usepackage{dcolumn}
\usepackage{bm}
\usepackage{tabularx,booktabs}
\usepackage{multirow}
\usepackage{float}
\usepackage[normalem]{ulem}

\begin{document}
	
	
	\title{{Bayesian tit-for-tat fosters cooperation in evolutionary stochastic games}}

	\author{Arunava Patra}
	\email{arunava20@iitk.ac.in}
	\address{
		Department of Physics,
		Indian Institute of Technology Kanpur,
		Uttar Pradesh 208016, India
	}
	
	\author{Supratim Sengupta}
	\email{supratim.sen@iiserkol.ac.in }
	\address{
		Department of Physical Sciences,
			Indian Institute of Science Education and Research Kolkata, Mohanpur Campus, West Bengal 741246, India
	}
	
	\author{Sagar Chakraborty}
	\email{sagarc@iitk.ac.in}
	\address{
		Department of Physics,
		Indian Institute of Technology Kanpur,
		Uttar Pradesh 208016, India
	}
	\begin{abstract}
		Learning from experience is a key feature of decision-making in cognitively complex organisms. Strategic interactions involving Bayesian inferential strategies can enable us to better understand how evolving individual choices to be altruistic or selfish can affect collective outcomes in social dilemmas. Bayesian strategies are distinguished, from their reactive opponents, in their ability to modulate their actions in the light of new evidence. We investigate whether such strategies can be resilient against reactive strategies when actions not only determine the immediate payoff but can affect future payoffs by changing the state of the environment. We use stochastic games to mimic the change in environment in a manner that is conditioned on the players' actions. By considering three distinct rules governing transitions between a resource-rich and a resource-poor states, we ascertain the conditions under which Bayesian tit-for-tat strategy can resist being invaded by reactive strategies. We find that the Bayesian strategy is resilient against a large class of reactive strategies and is more effective in fostering cooperation leading to sustenance of the resource-rich state. However, the extent of success of the Bayesian strategies depends on the other strategies in the pool and the rule governing transition between the two different resource states.
	\end{abstract}
	\maketitle
	
\section{Introduction}
Any organism infers, learns, and decides: Mere mechanical reaction to her opponent's actions and/or her changing environment's states repeatedly is not a realistic proposition in any social context. It is very natural that over time an organism would like to understand her opponent and environment better and better so as to adapt to the situation in a way most beneficial to her. Thus, it is not desirable to ignore the learning aspect of the faculty of mind~\cite{Yoshida2008} of the organism when trying to comprehend her decisions adopted with a view to seeking individual gains, either selfishly or altruistically. Needless to say, in a population of such strategically interacting organisms, the rise and sustenance of complex emergent phenomena like cooperation~\cite{Axelrod1981} and avoidance of the tragedy of the commons~\cite{Hardin1968, Ostrom1999}, would very much depend on the learning and inferring abilities of the organisms.

Cooperation, an act that benefits others at a personal cost, is a ubiquitous phenomenon in nature, observed across various biological~\cite{ Dugatkin1997, Connor1999, Bshary2006,Patzelt2014, Duguid2020, Loukola2024} and social systems~\cite{Shehory1999, Dasgupta2009, Enke2019} ranging from microbes to human societies. It therefore poses the challenge of explaining how cooperation is so widespread even though it appears to be incompatible with Darwinian evolution~\cite{Darwin1859}, which favors traits that maximizes an individual's fitness. A large body of work~\cite{Nowak1992, Brauchli1999, Santos2005, Ohtsuki2006, Milinski2002,Szolnoki2010, Santos2016, Basak2024, Sun2025} has been carried out to understand the key mechanism that promotes cooperation. Especially in a scenario of fluctuating resource-states, a stochastic game is a useful model for understanding the evolution of cooperation~\cite{Hilbe2018Nature, Kleshnina2023, Mondal2024SG}. 

Stochastic games \cite{Shapley1953, Hilbe2018Nature} were introduced to account for the possibility that individual decisions, whether altruistic or selfish, can change the state of the environment which in turn can influence further decision-making. In the context of benefits gained from a public resource, this reflects the fact that increased cooperation can enhance the benefits that accrue from the resource while increased defection can lead to lower benefits due to depletion of the resource. In a common version of a stochastic game, such changes are manifest through changes in the payoff structure that are conditioned on actions of the players. We, therefore, imagine a scenario where the players can switch randomly between two games differentiated by different benefits of mutual cooperation. These two games can reflect different resource states of the environment, one of which is more beneficial than the other. The transitions between these two resource (game) states is determined by the actions of the two players. For deterministic transition vectors, the action-dependent transition from one game state to the other occurs with probability 0 or 1. This allows for 64 possible transition vectors assuming that we do not distinguish between which of the two players cooperate when the other defects. 

Previous work has demonstrated \cite{Hilbe2018Nature} that stochastic switching between different resource states can lead to enhanced propensity for cooperation when compared to the outcome from a single game. The focus, in all these papers, was on exploring stochastic game dynamics using reactive or memory-one strategies in the context of a repeated games. Strategies were updated based on cumulative payoff received from interactions using the pairwise-comparison rule. Such frameworks incorporate a very restricted form of learning, via strategy updates, that is based on optimizing immediate payoffs. Moreover, it requires knowledge of the memory-n strategies of other players in the population. Such knowledge is often unavailable or hard to acquire. Learning to tune one's own strategies by using accumulated evidence collected from the opponent's actions over time is a hallmark of Bayesian updating~\cite{Bayes1763, Jaynes2003}. There are several instances where Bayesian update has been found to be successful in predicting behavior in both human~\cite{Krding2004,Langlois2025} and animal societies~\cite{McNamara2006,PrezEscudero2011}, although systematic deviations from optimal Bayesian updating, attributed to specific cognitive biases, have also been widely reported. 

In a recent work \cite{PatraNJP2024}, we investigated the efficacy of a Bayesian inferential strategy playing against a reactive strategy and found that the Bayesian strategy can outperform many reactive strategies. Given the sociological significance of switching between different resource states, it is pertinent to ask how a Bayesian inferential strategy would perform in such a setting. In this paper, we adopt the framework of stochastic games to examine the effectiveness of a Bayesian strategy in fostering cooperation. In the process, we develop the mathematical framework for studying the interaction between Bayesian and reactive strategies in stochastic games. We use two distinct donation games to represent the resource rich and resource poor states and analyze the effectiveness of a Bayesian strategy for three distinct transition vectors that determine the switching between the resource states in a manner that depends on the actions of the two players. 

Our results indicate that Bayesian inferential strategies can avoid being invaded by a large set of reactive strategies, particularly those that tend to be more selfish in the game with higher benefit for cooperation. However, the extent to which the Bayesian strategy is successful depends on the transition rule that controls switching between the two resource states. Moreover, in an evolutionary setting, a Bayesian strategy can enhance cooperation levels, dominate over other strategies and also ensure higher propensity to be in the more beneficial state, in the absence of Tit-for-Tat (TFT) strategies~\cite{Axelrod1981}. Our work reinforces the importance of Bayesian inferential strategies in fostering cooperative behavior in social dilemmas.


\section{Framework}
\label{sec:II}
\subsection{Overview}
We consider a two-player, two-action stochastic game in which a player can adopt either a reactive strategy or a Bayesian strategy (to be elaborated in Sec.~\ref{sec:sec_BS}) to interact repeatedly with her opponent. Cooperation ($C$) and defection ($D$) are the two possible actions in the underlying game. A pair of interacting partners can switch between two distinct resource states that are distinguished by differences in benefits associated with mutual cooperation, while the cost of cooperation remains the same in both resource states.  
To simplify our framework, we assume two resource (game) states ($s^1$) and ($s^2$), where the state $s^1$ yields a larger benefit for {mutually} altruistic behavior than $s^2$. 

In each resource state, we consider the underlying game to be a prisoner's dilemma (PD)~\cite{1965_RC}. For convenience, we write the payoff matrices using two parameters, $r_1$ and $r_2$ which correspond to the benefit-to-cost ratio in the beneficial state $s^1$, and the depleted state $s^2$, respectively. The parameters $r_1$ and $r_2$ should satisfy the condition $r_1>r_2$ as we have assumed that the $s^1$ is more profitable than $s^2$. The effective payoff matrices for the games in $s^1$ and $s^2$ are, respectively, given by:

\begin{subequations} 
	\begin{eqnarray}
		\label{eqn:non_getnp1}
		{\sf U_1}\equiv
		\begin{blockarray}{ccc}
			&  C & D &  \\
			\begin{block}{c[cc]}
				C~ & r_1-1 & -1\\
				D~ & r_1 & 0\\ 
			\end{block}
		\end{blockarray}~,  \\	
		\label{eqn:non_gentq1}
		{\sf U_2}\equiv
		\begin{blockarray}{ccc}
			&  C & D &  \\
			\begin{block}{c[cc]}
				C~ & r_2-1 & -1\\
				D~ & r_2 & 0\\ 
			\end{block}
		\end{blockarray}~.	
	\end{eqnarray}
\end{subequations}
Each benefit-to-cost ratio parameter must satisfy the following two conditions: $r_i>r_i-1>0>-1$ and $2(r_i-1)>(r_i-1)$, which are required for the game to be a repeated PD game.

While repeated game between reactive players is long-established, treatment of such a game in the presence of a Bayesian player is rather new in the literature~\cite{PatraNJP2024, KleimanWeiner2025}. 
Since the framework is fairly technical, we first present an overview of the key ideas of the Bayesian play in simple terms while deferring the technicalities to the subsequent sections.
\begin{figure*}[t]
	\centering
	\includegraphics[width=\textwidth]{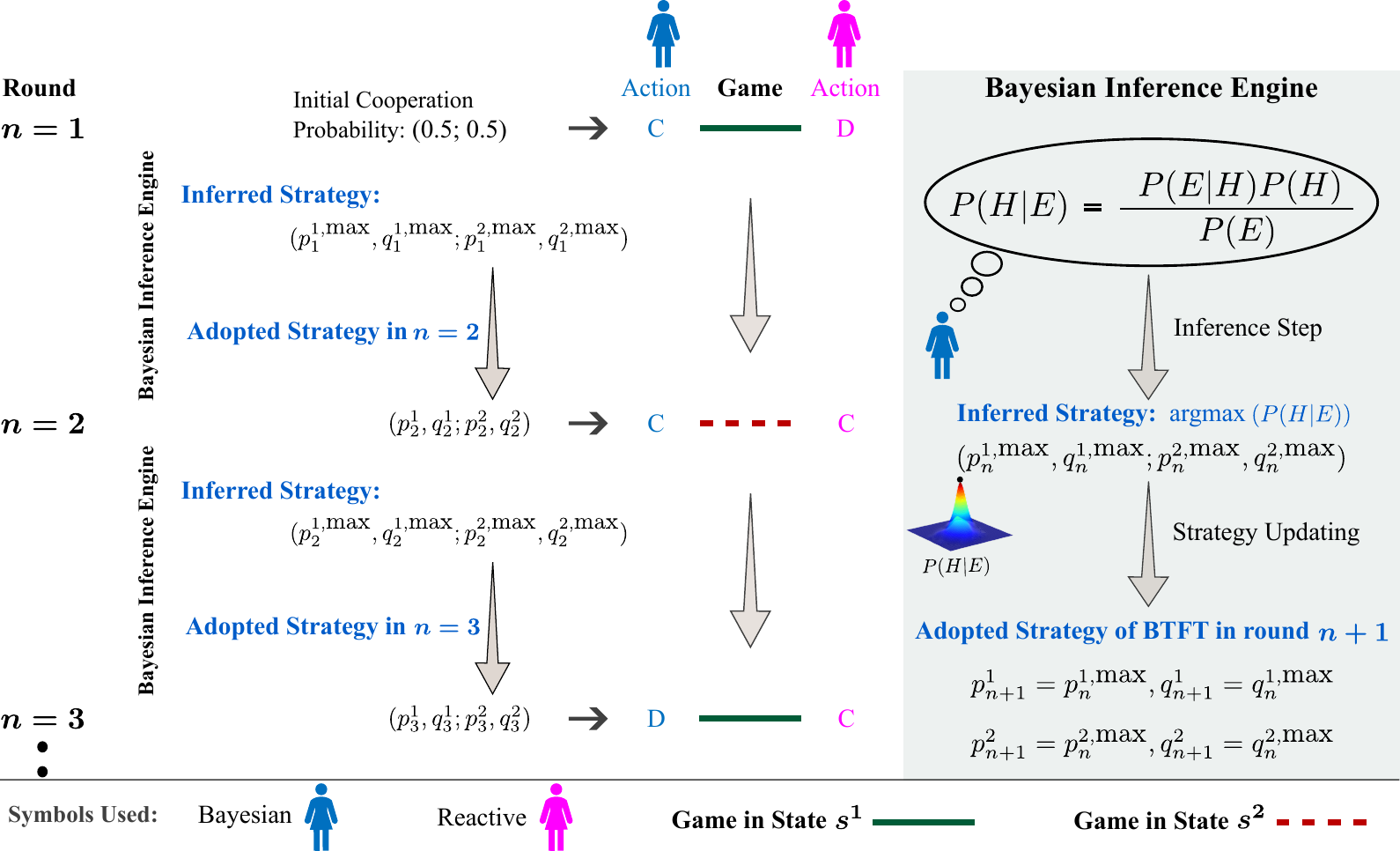}	
	\caption{ A schematic figure showing the interaction between a Bayesian player (shown in blue) and a reactive player (shown in pink) over time. The interaction between the two can occur in game state $s^1$ (represented by a solid green line) or state $s^2$ (represented by a dashed red line). We assume that all interactions initially start in state $s^1$ and the transition vector is $\tau=(1,0,0,0,1,0,0,0)$ such that only mutual cooperation in state $s^1$ ensures that both players remain in state $s^1$ and only mutual cooperation in state $s^2$ leads to a switch from state $s^2$ to state $s^1$. The Bayesian player uses a Bayesian inference engine (depicted in the right half of the figure) to infer the strategy of her opponent at the end of each round and then adopts the inferred strategy as her own strategy in the subsequent round. The inference engine starts with a prior hypothesis $(H)$ about the opponent's strategy and uses the actions of the opponent as evidence $(E)$ to continuously update her hypothesis over time following Bayes' rule.} 
	\label{fig:schematic}
\end{figure*}%

Fig.~\ref{fig:schematic} is a schematic diagram that showcases how the interaction between a Bayesian player and her reactive opponent unfolds over time. When a Bayesian player interacts with a reactive opponent, the former can use the actions of the reactive opponent to dynamically infer the opponent's strategy. We assume that the game starts in state $s^1$. In the first round, the Bayesian player is indifferent about the action to be played in both states and cooperates with a probability 0.5 in each state in accordance with principle of insufficient reason. The reactive player takes an action based on her reactive strategy. These actions can change the state of the game (depending on the transition rule governing transitions between the different resource states) in the second round, as shown by the red dashed line in Fig.~\ref{fig:schematic}. These sequence of play proceeds \emph{ad infinitum}.

We can think of the Bayesian player as one possessing a Bayesian inference engine in her brain that allows her to use the opponent's action as evidence to revise her belief about the opponent's reactive strategy. The inference engine computes the posterior probability distribution, given the prior and the likelihood (see Sec.~\ref{sec:sec_BS} for details) and determines the maxima of the posterior distribution which is then used as her inferred belief about the opponent's reactive strategy at the end of the round. After the Bayesian player has updated her {\it belief} about the opponent's strategy in a given round, she has to select an action on the basis of her new belief. Among the many possibilities of selecting an action, we focus on a Bayesian strategy that is based on the concept of probability matching where the Bayesian player adopts the inferred belief about the opponent's strategy as her own strategy in the subsequent round. Following \cite{PatraNJP2024}, we call such a strategy Bayesian Tit-for-Tat (BTFT).

\subsection{Reactive strategy and transition rules}
Let us assume that the focal  player uses a {fixed} reactive strategy which we denote using the symbol $S^r$. If we let $p^{i(r)}$ and $q^{i(r)}$ represent the focal player's cooperation probability in the state $s^i$ conditioned on the opponent's action (C and D, respectively) in the last round, then $S^r$ is mathematically parametrized by four parameters, each ranging between zero to one:($p^{1(r)}$, $q^{1(r)}$; $p^{2(r)}$, $q^{2(r)}$).
We emphasize that these probabilities are {\it not} conditioned on the resource state in the last round and are employed regardless of the resource state in which both player's were in the last round. As an illustration, we represent the pure/corner strategies using the above notation: (ALLD; ALLD) is represented by $(0,0; 0,0)$; (ALLC; ALLC) corresponds to $(1,1; 1,1)$; $(1,0;1,1)$ represents (TFT; ALLC); (TFT; TFT) corresponds to $(1,0;1,0)$; and $(1,0;0,0)$ denotes the pure strategy (TFT; ALLD). Here, ALLC, ALLD, and TFT are the standard abbreviations in the theory of repeated games and correspond to `always cooperate', `always defect', and `tit-for-tat' respectively.

The transition between these resource states is determined by the actions of the players. We define the transition rule {by specifying the transition vector} $\bm{\tau}=\left(\tau^1_{CC}, \tau^1_{CD}, \tau^1_{DC},\tau^1_{DD}; \tau^2_{CC},\tau^2_{CD},\tau^2_{DC}, \tau^2_{DD}\right)$, where $\tau^i_{a\tilde{a}}$ is the transition probability from the state $i$ to the more profitable state ($s^1$) if the actions of focal player and her opponent players are $a\in\{C, D\}$ and $\tilde{a}\in\{C, D\}$ respectively. Hence, the superscript $1$ in the components of the transition probability vector denotes the probability of remaining in the more profitable state $s^1$ while the superscript $2$ denotes the transition probability from $s^{2}\rightarrow s^{1}$. 

For simplicity, we choose deterministic transition vectors such that each element $\tau^i_{a\tilde{a}}=\{0,1\}$. We assume that the transition probabilities depend only on the number of cooperators and defectors in an interaction but doesn't depend on who cooperated or defected {\it i.e.} $\tau^i_{CD}=\tau^i_{DC}$ $\forall i$. Therefore, there are $2^6$ possible transition vectors. However, we select a sub-set of 3 transition vectors by tuning the condition for transition back from a depleted state to the more beneficial state, while keeping the condition for transition to the depleted state from the more beneficial state unchanged. We assume that any defection in the beneficial state $(s^1)$ leads to a transition to the game state $(s^2)$ that offer lower benefits for mutual cooperation i.e., $\tau^1_{CD}=\tau^1_{DD}=0$. In contrast, mutual cooperation in the more beneficial state ensures that the system remains in that state. Thus, the possible transition vectors take the form $\left(1, 0, 0; \tau^2_{CC},\tau^2_{CD},\tau^2_{DD}\right)$. If we then consider three distinct scenarios for return to the more beneficial state ($s^1$) from the depleted state ($s^2$): (i) only mutual cooperation in the depleted state can ensure return to the more beneficial state, (ii) cooperation by at least one of the two players ensures return to the more beneficial state, (iii) transition to game state $s^1$ occurs regardless of the action of the two players, we are left with three possible transition vectors: ${\bm \tau_{00}}=(1,0,0;1,0,0)$, ${\bm \tau_{10}}=(1,0,0;1,1,0)$ (called a time-out game with conditional return) and ${\bm \tau_{11}}=(1,0,0;1,1,1)$ (called a time-out game \cite{Hilbe2018Nature}).

For a more intuitive interpretation of the three transition vectors, it helps to view the system through the lens of the tragedy of the commons (ToC)~\cite{Hardin1968, Ostrom1999, Weitz2016,Tilman2020}---the selfish over-exploitation of a common resource. Considering the common resource between two players, obviously, the ToC is strongest and weakest when the transition is driven by ${\bm \tau_{00}}$ and ${\bm \tau_{11}}$ respectively. This is because any defection leads to the depleted state for ${\bm \tau_{00}}$; in contrast, for ${\bm \tau_{11}}$, the resource state is always altered from a depleted to a beneficial state irrespective of the players' action in the depleted state. 
ToC is intermediate when the transition vector is ${\bm \tau_{10}}$. Because the defection of both players causes degradation, this game is called a time-out game with a conditional return. The resource in the game corresponds to ${\bm \tau_{10}}$ and ${\bm \tau_{11}}$ could be associated with self-renewing resources, as the defection in the depleted state could not deplete further. Resource renews in the deplete state more than in the replete state. However, the resource in the game corresponds to ${\bm \tau_{00}}$ is not self-renewing as only cooperators enrich the resource, and defection keeps both players in the resource depleted state.


\subsection{ A Bayesian player in a stochastic game}\label{sec:sec_BS}


A player employing a Bayesian strategy tries to infer the opponent's strategy. For instance, if an opponent has reactive strategy, $\tilde{S}^r$, parametrized as ($\tilde{p}^{1(r)}$, $\tilde{q}^{1(r)}$; $\tilde{p}^{2(r)}$, $\tilde{q}^{2(r)}$), then the Bayesian focal player would try to infer $\tilde{S}^r$ round-by-round; and then modulate her own action in each round on the basis of round-wise inferences. (Note that here and henceforth we use $\sim$ (tilde) over symbols to specify the opponent of the focal player.)

A complete specification of $S^b$ requires specifying a set of four parameters (each lying between zero and one) that changes over rounds ($n$): ($p^{1(b)}(n)$, $q^{1(b)}(n)$; $p^{2(b)}(n)$, $q^{2(b)}(n)$).  Here, $p^{i(b)}(n)$ and $q^{i(b)}(n)$ represent the Bayesian player's cooperation probability in the state $s^i$ conditioned on the opponent's action ($C$ and $D$, respectively) in the $(n-1)$th round. 
The Bayesian player continuously updates her {\it belief} about the opponent's strategy on the basis of observed evidence collected from the opponent's actions ($C$ or $D$) in a specific resource state over multiple rounds. Formally, this evidence at $n$th round is represented by $E_n\in \{(s^1,C), (s^1,D), (s^2,C), (s^2,D)\}$, where each pair $({s}^i, \tilde{a})$ denotes the observed state $s^i \in \{s^1, s^2\}$ and the opponent’s action $\tilde{a} \in \{C, D\}$.  This updating process follows the Bayes' rule, which requires a prior probability distribution of the possible strategies that can be adopted by her opponent. We consider the initial prior distribution $P_{1}(p^1,q^1; p^2,q^2 )$ to be a {\it uniform} distribution over all possible strategies. The Bayesian player updates her {\it belief} $(p^1,q^1; p^2,q^2)\in[0,1]\times[0,1]\times[0,1]\times[0,1]$ about the opponent's reactive strategy at the end of round $n$ from the posterior probability distribution,
\begin{equation}
	P_{n}(p^1,q^1; p^2,q^2|E_n)=\frac{P(E_{n}|p^1,q^1; p^2,q^2)P_{n}(p^1,q^1; p^2,q^2)}{P(E_n)},\label{eq:Bayes}
\end{equation}
where $P(E_n)=\sum\limits_{p^1=0}^1 \sum\limits_{q^1=0}^1\sum\limits_{p^2=0}^1 \sum\limits_{q^2=0}^1 P(E_{n}|p^1,q^1; p^2,q^2)$ $P_{n}(p^1,q^1; p^2,q^2)$ is the marginal likelihood or model evidence.
The subscript $n$ indicates that the corresponding quantities are evaluated in the $n^{th}$ round.

The belief updating process is recursive: the prior distribution in the $(n+1)^{th}$ round is the posterior distribution found at the end of the $n^{th}$ round, i.e., $P_{n+1}(p^1, q^1; p^2, q^2) = P_{n}(p^1, q^1; p^2, q^2|E_n)$. At the end of the $n^{th}$ round, the BTFT player identifies the arugument $(p^{1, \rm max}_{n}, q^{1, \rm max}_{n}; p^{2, \rm max}_{n}, q^{2, \rm max}_{n})$ at which the posterior distribution $P_n(p^1, q^1; p^2, q^2 \mid E_n)$ achieves its global maximum. This argument serves as her revised inferred strategy in the $(n+1)^{th}$ round {\it i.e.} $S^b$:($p^{1(b)}(n+1)=p^{1, \rm max}_{n}$, $q^{1(b)}(n+1)=q^{1, \rm max}_{n}$; $p^{2(b)}(n+1)=p^{2, \rm max}_{n}$, $q^{2(b)}(n+1)=q^{2, \rm max}_{n}$). 
In cases where the posterior distribution exhibits multiple maxima, the Bayesian player randomly selects one of them. Henceforth, $S^b$ will denote BTFT strategy.

{In passing, we would like to explain our reason for adopting the name BTFT for the Bayesian strategy described above. In fact, it closely follows the spirit of TFT nomenclature: The BTFT player believes that the posterior global maximum corresponds to the opponent's true strategy played in the previous round, since it is the most probable under the posterior; thus, by adopting the inferred strategy, she could \emph{mimic} the opponent's true strategy. In a way, analogous to what happens in TFT, she is \emph{reciprocating} with the strategy played by her opponent---only difference being in TFT, actions ($C$ and $D$) are mimicked; whereas in BTFT, strategies ($p$'s and $q$'s) are mimicked. It is worth emphasizing that this line of thinking---the most probable strategy could be the opponent's true strategy---occurs only in the Bayesian player's mind: The inferred strategy may not coincide with the strategy actually employed by the reactive opponent.}

To do further numerical calculations, one would need explicit form of the likelihood used in Eq.~(\ref{eq:Bayes}). As detailed in Appendix~\ref{app:lhf}, $P(E_{n}|p^1,q^1; p^2,q^2)$ naturally has following form:
\begin{subequations}
	
	\begin{eqnarray}
&&		P(E_{n}|p^1,q^1; p^2,q^2)=\nonumber\\ &&p^1\{\tau^1_{CC}\sigma^1_{CC, n-1} +\tau^1_{DC}\sigma^1_{DC, n-1}\} +\nonumber\\ &&q^1\{\tau^1_{CD}\sigma^1_{CD,n-1}+\tau^1_{DD}\sigma^1_{DD, n-1}\}+\nonumber\\
		&&p^1\{\tau^2_{CC}\sigma^2_{CC, n-1} +\tau^2_{DC}\sigma^2_{DC, n-1}\}+\nonumber\\
		&&q^1\{\tau^2_{CD}\sigma^2_{CD, n-1}+\tau^2_{DD}\sigma^2_{DD, n-1}\} \nonumber\\
		&&~~~~~~~~~~~~~~~~~\text{if}~~E_n= (s^1, C),\label{eq:s1C}
	\end{eqnarray}
	
	\begin{eqnarray}
		&& P(E_{n}|p^1,q^1;p^2,q^2)=\nonumber\\ &&(1-p^1)\{\tau^1_{CC}\sigma^1_{CC, n-1} +\tau^1_{DC}\sigma^1_{DC, n-1}\} +\nonumber\\ &&(1-q^1)\{\tau^1_{CD}\sigma^1_{CD,n-1}+\tau^1_{DD}\sigma^1_{DD, n-1}\}+\nonumber\\
		&&(1-p^1)\{\tau^2_{CC}\sigma^2_{CC, n-1} +\tau^2_{DC}\sigma^2_{DC, n-1}\}+\nonumber\\
		&&(1-q^1)\{\tau^2_{CD}\sigma^2_{CD, n-1}+\tau^2_{DD}\sigma^2_{DD, n-1}\} \nonumber\\
		&&~~~~~~~~~~~~~~~~~~~~~~~~~\text{if}~~E_n= (s^1, D),
	\end{eqnarray}
	
	\begin{eqnarray}
		&&P(E_{n}|p^1,q^1; p^2,q^2)=\nonumber\\ &&p^2\{(1-\tau^1_{CC})\sigma^1_{CC, n-1} +(1-\tau^1_{DC})\sigma^1_{DC, n-1}\} +\nonumber\\ &&q^2\{(1-\tau^1_{CD})\sigma^1_{CD, n-1}+(1-\tau^1_{DD})\sigma^1_{DD, n-1}\}+\nonumber\\
		&&p^2\{(1-\tau^2_{CC})\sigma^2_{CC, n-1} +(1-\tau^2_{DC})\sigma^2_{DC, n-1}\}+\nonumber\\
		&&q^2\{(1-\tau^2_{CD})\sigma^2_{CD, n-1}+(1-\tau^2_{DD})\sigma^2_{DD, n-1}\} \nonumber\\
		&&~~~~~~~~~~~~~~~~~~~~~~~~~~~~~~~~~~\text{if}~~E_n= (s^2, C),
	\end{eqnarray}
	
	\begin{eqnarray}
		&&P(E_{n}|p^1,q^1; p^2,q^2)=\nonumber\\ &&(1-p^2)\{(1-\tau^1_{CC})\sigma^1_{CC, n-1} +(1-\tau^1_{DC})\sigma^1_{DC, n-1}\} +\nonumber\\ &&(1-q^2)\{(1-\tau^1_{CD})\sigma^1_{CD, n-1}+(1-\tau^1_{DD})\sigma^1_{DD, n-1}\}+\nonumber\\
		&&(1-p^2)\{(1-\tau^2_{CC})\sigma^2_{CC, n-1} +(1-\tau^2_{DC})\sigma^2_{DC, n-1}\}+\nonumber\\
		&&(1-q^2)\{(1-\tau^2_{CD})\sigma^2_{CD, n-1}+(1-\tau^2_{DD})\sigma^2_{DD, n-1}\} \nonumber\\
		&&~~~~~~~~~~~~~~~~~~~~~~~~~~~~~~~~~~~~~~\text{if}~~E_n= (s^2, D). 
	\end{eqnarray}
	\label{eq:likelihood_gen}
\end{subequations}

Here $\sigma^i_{a\tilde{a}, n}$ denotes the component of the state vector ${\bm \sigma}_n$ and represents the unconditional probability that the focal player and opponent's action are $a,\tilde{a}$ respectively, in state $s^i$ during the $n^{\text{th}}$ round. The quantities $p^1$, $q^1$, $p^2$, and $q^2$ are probabilities sampled from the interval $[0,1]$. This choice of likelihood function is a generalization of the likelihood function when there is just one game state \cite{PatraNJP2024}. Such a choice is appropriate because a reactive player characterized by $p^{1(r)} = p^1$, $q^{1(r)} = q^1$, $p^{2(r)} = p^2$, and $q^{2(r)} = q^2$ selects action $\tilde{a}\in \{C, D\}$ in state $s^i \in \{s^1, s^2\}$ in the $n^{\text{th}}$ round with probability given by Eq.~(\ref{eq:likelihood_gen}), assuming the state vector in the preceding round is ${\bm \sigma}_{n-1}$. 

For the estimation of the likelihood function at any $n$, we need to know the initial value ${\bm \sigma}_{1}$ when $n=1$, which can be calculated from the initial probability of actions $a$ and $\tilde{a}$ of both the focal Bayesian and its opponent as follows: We assume that the repeated stochastic game can start in any of the two possible resource states with equal probability of 0.5. Furthermore, we assume that both reactive and Bayesian players are indifferent to which action is played in the initial round---consequently, in line with the principle of insufficient reason, each player cooperates with a probability of $0.5$ in the initial round. This implies that the probability of any action pair in any game state in the initial round is given by $0.5\times 0.5\times 0.5=0.125$, i.e., $\sigma^i_{a\tilde{a}, 1}=0.125, \forall a, \tilde{a}\in \{C, D\}$ and $\forall i\in\{1,2\}$. This initial state vector has also been used for corner reactive strategies (cf. Appendix~\ref{app: true_corner})---$p^i,q^i\in\{0,1\}$.

Finally, we note that the inferred strategy of the Bayesian player keeps evolving over time until the belief converges and the Bayesian player is able to infer the true reactive strategy of her opponent. In the case where the opponent is also a Bayesian player, the focal Bayesian player continues to update her strategy according to Bayes' rule; however in this case there cannot be any convergence of belief as the opponent does not have a fixed strategy. The numerical simulation of the inference process and the evaluation of the payoff of BTFT are provided in Appendix~\ref{app:finding_Payoff matrix_numerically}.

\section{Questions}
We are essentially interested in the evolutionary robustness of BTFT players in a population of reactive players. We would like to ascertain the potential evolutionary advantage of BTFT players in both replication-selection~\cite{Traulsen2006_PRE,Nowak2006Book} and mutation-selection regimes~\cite{Fudenberg2006_JET, Imhof2009}. As a result, we ask the following two central questions of this paper. 

\subsection{Is BTFT evolutionarily stable?}
\label{sec:sec_ES_BTFT}
It is worthwhile to ask at this point how a Bayesian strategy fares against the reactive opponent. To address this question, we consider a well-mixed population of reactive and Bayesian players where a focal player can randomly meet with either a Bayesian or a reactive player. Therefore, the population has three distinct types of interaction: reactive-reactive, reactive-Bayesian, and Bayesian-Bayesian. Hence, the payoff matrix of the focal player corresponding to two players repeatedly interacting can be represented as
\begin{equation}	
	{\sf\Pi}\equiv~
	\begin{blockarray}{ccc}
		&  \textrm{Reactive} & \textrm{Bayesian} &  \\
		\begin{block}{c[cc]}
			\textrm{Reactive}~ & \pi(S^r,{S}^r) & \pi(S^r,{S}^b)\\
			\textrm{Bayesian}~ & \pi(S^b,{S}^r) & \pi(S^b,{S}^b)\\ 
		\end{block}
	\end{blockarray}
	\label{eq:Pi}
\end{equation}%
to model the strategic competition between reactive and Bayesian strategies.

During the calculation of the payoff elements in payoff matrix in Eq.~(\ref{eq:Pi}), we consider for generality that the payoffs are discounted by a discount factor $\delta\in(0,1)$. The discount factor can be equivalently interpreted as the probability that the game is played in the subsequent round. For example, the probability that the $n$th round occurs is given by $\delta^{n-1}$. Therefore, a focal player with strategy $S$ interacting with opponent of strategy $\tilde{S}$ gets on average following payoff per round in the repeated game:
\begin{equation}
	\pi({S}, \tilde{ S})=(1-\delta)\sum_{n=1}^{n=n_f} u_n\delta^{n-1}.
	\label{eq:Average_payoff_calculation}
\end{equation}
Here, $u_n$ $\in$ $\{r_1,-1,r_1-1,0\}\cup \{r_2,-1,r_2-1,0\}$ (see Eq.~	\ref{eqn:non_getnp1} and Eq.~\ref{eqn:non_gentq1}) is the payoff of focal player in $n$th round. The factor ($1-\delta$) comes because we have divided the total accumulated payoff at the end of repeated game by the expected number of rounds (also known as effective game length) which is given by $\sum_{n=1}^{\infty}\delta^{n-1}=\frac{1}{1-\delta}$, assuming $n_f=\infty$. Since the sequence $(u_n)_{n\in\mathbb{N}}$ is a random sequence, one must work with an average value of ${\sf \Pi}$ calculated in a number of independent trials. The payoff elements can be evaluated numerically: Details, especially when BTFT is involved, are provided in Appendix~\ref{app:finding_Payoff matrix_numerically}. 

The standard theory of evolutionary games tells us that the evolutionary stability~\cite{Smith1973,MaynardSmith1974} of a Bayesian strategy relative to any arbitrary reactive strategy in an infinitely large population can be determined by comparing the payoffs in Eq.~(\ref{eq:Pi}). BTFT is an evolutionarily stable strategy (ESS) if $\pi(S^b, S^b)>\pi(S^r, S^b)$ or if $\pi(S^b, S^b)=\pi(S^r, S^b)$ and $\pi(S^b, S^r)>\pi(S^r, S^r)$. The reactive strategy is an ESS if  $\pi(S^r, S^r)>\pi(S^b, S^r)$ or if $\pi(S^r, S^r)=\pi(S^b, S^r)$ and $\pi(S^r, S^b)>\pi(S^b, S^b)$. Both the strategies are ESS if $\pi(S^b, S^b)>\pi(S^r, S^b)$ and $\pi(S^r, S^r)>\pi(S^b, S^r)$, whereas neither is an ESS if $\pi(S^b, S^b)<\pi(S^r, S^b)$ and $\pi(S^r, S^r)<\pi(S^b, S^r)$. The last case corresponds to the existence of a stable mixed-state equilibrium between both Bayesian and reactive strategies and is referred to as a mixed ESS.

We wish to find out that under what conditions, if at all, BTFT can be ESS so as to ensure that any invasion by a mutant reactive strategy is effectively countered by the host population of Bayesian players.

\subsection{Does BTFT foster cooperation?}
\label{sec:evo_dynamics}
Next consider a well-mixed population of finite size $N$ where initially all the individuals in the population have the same strategy, i.e., the population is monomorphic. Since we are interested in the emergence and sustainability of cooperation, we assume that all the individuals initially use the ALLD strategy irrespective of the resource state. We wish to understand how this initial extremely selfish strategy can be displaced by a more cooperative one and ascertain the nature of those emergent strategies that can eventually resist invasion by other strategies. 

As is standard in the literature, in order to study such an evolutionary dynamics, we introduce new strategies into the population as rare mutants. The rare mutation limit means that a second mutant does not appear until the first mutant either invades the resident population completely or becomes extinct. If the mutant takes over the resident population, then the mutant becomes the new resident in the monomorphic population. Subsequently, another random mutant emerges, which either invades or goes extinct, and the mutation-selection process proceeds \emph{ad infinitum} in this manner. The chance of selection of a mutant is proportional to its fixation probability in the underlying stochastic replication-selection process: This selection protocol may be termed Imhof--Nowak--Fudenberg process following \cite{Fudenberg2006_JET, Imhof2009}~(see Appendix~\ref{app:INF_process}).  Note that in the transient state, the population has no more than two strategies at any given time, as the resident strategy and the mutant strategy always compete with each other; however, on a large time scale, the population is monomorphic, and transition occurs between the different monomorphic population states. 

Let the strategy set from which the mutants are drawn be denoted by $\mathcal{S}$: $\mathcal{S}$ includes the BTFT strategy $S^b$ and a number of reactive strategies $\{S^r\}$; an arbitrary element of $\mathcal{S}$ will be denoted by $S$ (with convenient subscripts or superscripts as and when needed). On average, in a round, the probability of witnessing action $C$ is a measure of how cooperative the population with strategy $S$ is. We denote this as $\gamma({S}, {S})$ and refer to it as the self-cooperation rate in the population. See  Appendix~\ref{app:cooperation_rate_reactive_strategy} and Appendix~\ref{app:cooperation_rate_BTFT} for detailed expressions. Next, in the Imhof--Nowak--Fudenberg process, which generates a temporal sequence of monomorphic populations, one could track what fraction, $x_S(t)$, of times population with strategy $S$ appears till time $t$. Hence, the \emph{average self-cooperation rate} at time $t$ is given by
\begin{equation}
	\langle \gamma(t)\rangle=\sum_{{S}\in\mathcal{S}}x_{S}(t)\gamma({S}, {S}).
	\label{eq:average_cooperation_markov_process}
\end{equation}
We are interested in knowing if BTFT strategy can enhance $\langle\gamma(t)\rangle$.

Moreover, since one would want to avoid the tragedy of the commons, we are also interested in finding out if the presence of the BTFT strategy enhances the probability of the environment being in the beneficial state, $s^1$. In more technical terms, analogous to the average self-cooperation rate, discussed above, we are interested to know if BTFT can elevate \emph{beneficial state level} defined as follows:
\begin{equation}
	\langle \alpha(t)\rangle=\sum_{{S}\in\mathcal{S}}x_{S}(t)\alpha({S}, {S}).
	\label{eq:benficial_state_prob_INF}
\end{equation}%
Here, $\alpha({S}, {S})$ is the probability of witnessing beneficial state $s^1$ on average in a round.


\section{ Results}
We use the framework described in the previous sections to set up extensive numerical simulations in order to address the questions raised in the preceding section. The relevant numerical codes employed to arrive at the results in this paper have been made publicly available on GitHub~\footnote{\url{https://github.com/ArunavaHub/BayesianStochasticGame.git}}. 
\subsection{Evolutionary stability of BTFT}
\begin{figure}
	\centering
	\includegraphics[scale=0.45]{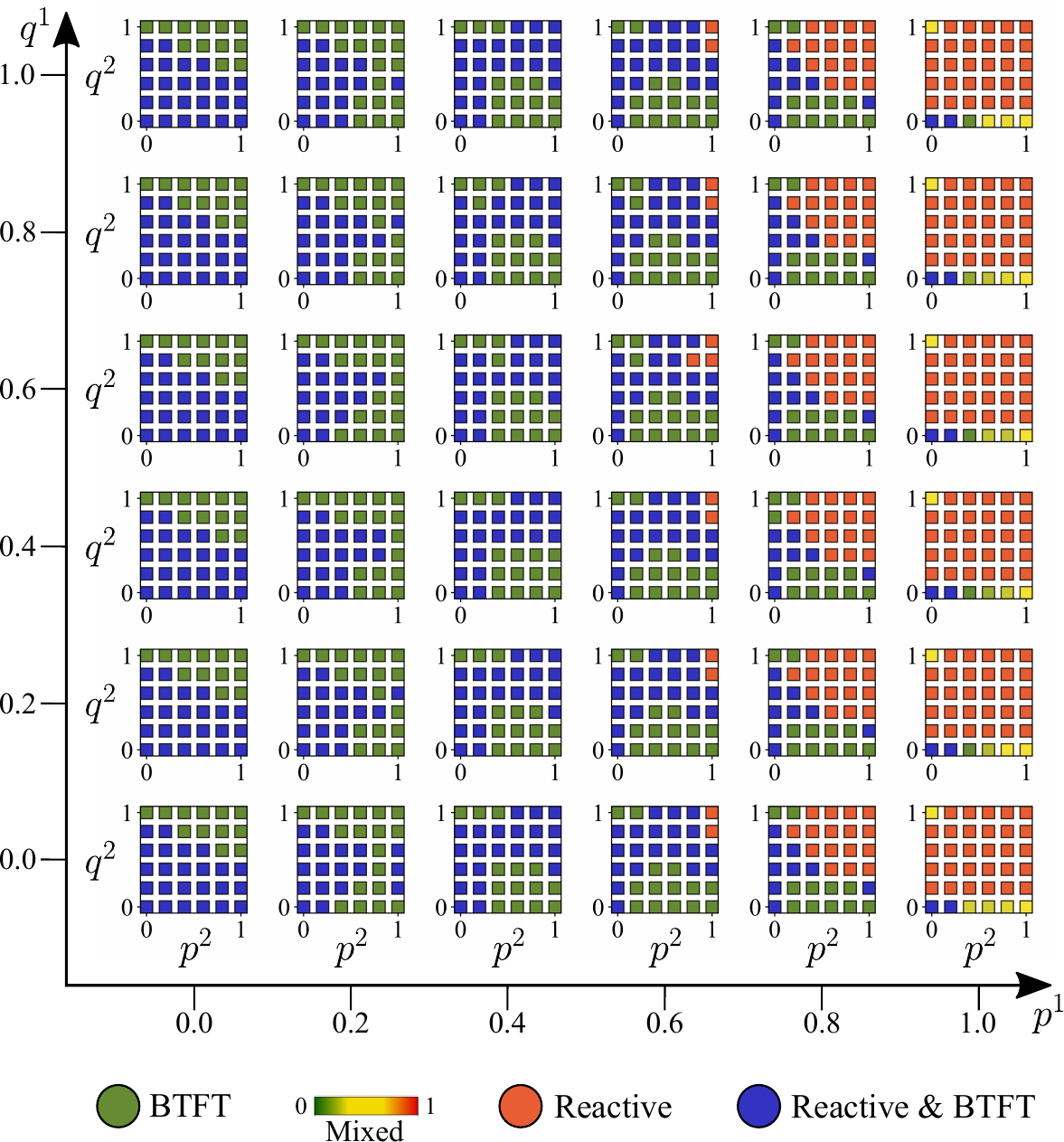}	
	\caption{ESS phase diagram for BTFT vs. reactive strategies in an infinite population for the transition vector ${\bm \tau_{00}}$.  The color gradient, ranging from 0 to 1, indicates the proportion of the reactive strategy in the mixed ESS. Blue corresponds to the scenario where both BTFT and the reactive strategy are an ESS; green indicates that only BTFT is an ESS; while red indicates that only the reactive strategy is an ESS.}
	\label{fig:ess_q00}
\end{figure}%
We examine here the evolutionary stability of the Bayesian strategy in a well-mixed, infinitely large population by determining the payoff matrices (see Eq.~\ref{eq:Pi}) when the Bayesian strategy is made to compete against various reactive strategies. 

\begin{figure}
	\centering
	\includegraphics[scale=0.45]{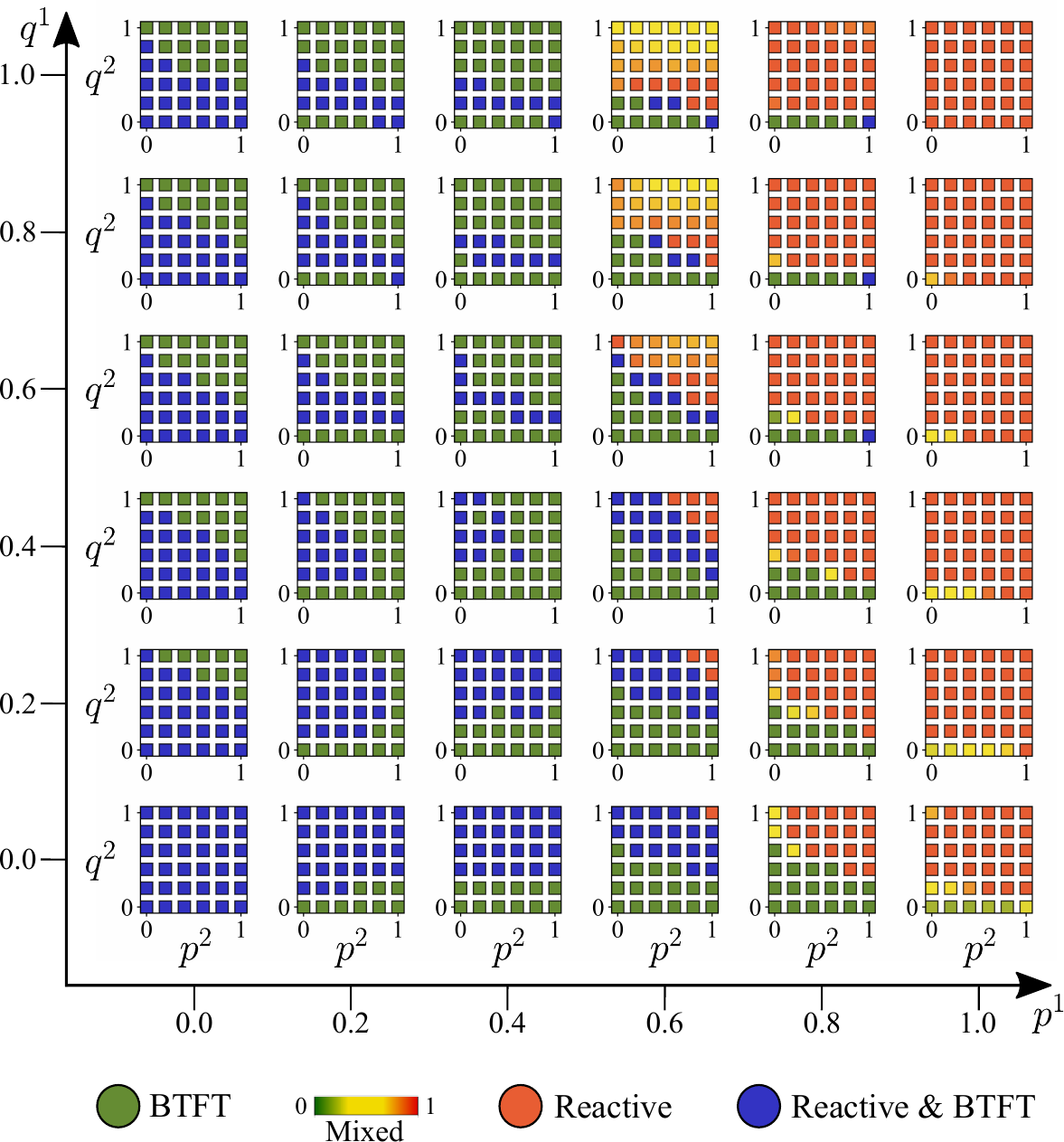}	
	\caption{ESS phase diagram for BTFT vs. reactive strategies in an infinite population for the transition vector ${\bm \tau_{10}}$. The color scheme followed is the same as in Fig.~\ref{fig:ess_q00}}
	\label{fig:ess_q10}
\end{figure}

\begin{figure}[h]
	\centering	
	\includegraphics[scale=0.45]{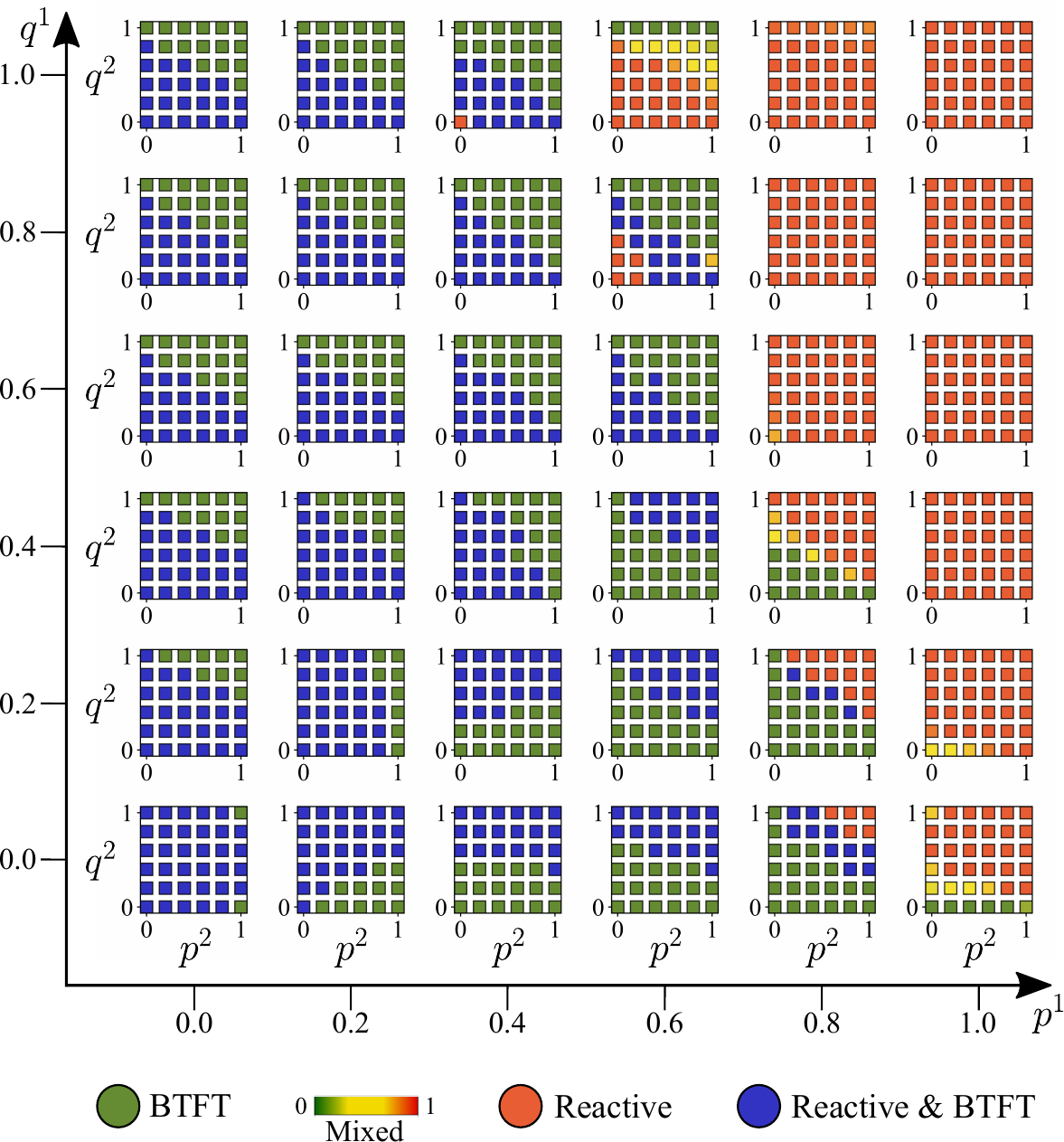}
	\caption{ESS phase diagram for BTFT vs. reactive strategies in infinite population for the transition vector ${\bm \tau_{11}}$. The color scheme followed is the same as in Fig.~\ref{fig:ess_q00}}
	\label{fig:ess_q11}
\end{figure}

We present our results using as ESS phase diagrams that pictorially represent the different outcomes of evolutionary stability analysis with different colors, across the entire space of reactive strategies. To make the ESS phase diagrams in Fig.~\ref{fig:ess_q00}, Fig.~\ref{fig:ess_q10} and Fig.~\ref{fig:ess_q11}, we partition the four dimensional reactive strategy space into $6^4$ grid points as we uniformly split the components of reactive strategy, $p^i$s and $q^i$s, into six grid points each. Then, at each grid point, we simulate repeated Prisoner’s Dilemma games of reactive versus reactive, reactive versus BTFT and BTFT versus BTFT to numerically find the payoff elements and construct the payoff matrix in Eq.~(\ref{eq:Pi}). We subsequently determine which strategy is ESS using this payoff matrix and the condition of evolutionary stability mentioned in Section~\ref{sec:sec_ES_BTFT}. Regions in the strategy space that correspond to the existence of a mixed state equilibrium where both the reactive and Bayesian strategy are present with non-zero fraction are represented using a color gradient. The frequency of the reactive strategy in the mixed-state equilibrium increases as the color gradient shifts towards red.

Since the BTFT needs to sample the opponent's action over a reasonable number of rounds to learn the opponent's strategy, the discount factor must be high so that the effective game length is high. Therefore, we make the ESS phase diagram for the discount factor $\delta=0.9$. From the phase diagram, it is evident that whether BTFT is an ESS depends on the nature of the reactive opponent's strategy and the transition vector. The benefit-to-cost ratios are fixed at $r_1=10$ and $r_2=2$ for concreteness.

For the transition vector ${\bm \tau_{00}}$, we see from Fig.~\ref{fig:ess_q00} that the BTFT is an ESS for a large fraction of reactive strategies in the reactive strategy space. Notably, the Bayesian strategy is particularly successful in preventing the invasion of those reactive mutants that are less likely to reciprocate the altruistic action of their opponent in the more beneficial game ($(p^1\leq0.4, q^1; p^2, q^2)~ \forall q^1, p^2, q^2\in[0,1]$). On the other hand, reactive residents having a strategy $(p^1=1, q^1; p^2, q^2)~ \forall q^1, p^2, q^2\in[0,1]$ except $p^2=0.0,0.2$ and $q^2=0$ inhibit the invasion by a Bayesian mutant. Reactive strategies start dominating over their Bayesian opponents as the value of $p^1$ of the reactive strategy increases from 0.4. This is manifest in Fig.~\ref{fig:ess_q00} by the proliferation the region where only the reactive strategy is an ESS (represented by red color). Reactive strategies that reciprocate, with high probability, the altruistic act of the opponent in the more beneficial game are more likely to invade a Bayesian strategy. This likelihood of invasion increases when the probabilities of cooperation ($p^2, q^2$) in the lower benefit game are moderately high as well. 
Among the mutants with pure reactive strategy, BTFT cannot prevent the invasion by the mutants (ALLC; ALLC) and (TFT; ALLC). However, it is apparent from Fig.~\ref{fig:ess_q00} that the BTFT can avoid being invaded by a purely selfish strategy like (ALLD; ALLD).

A similar pattern is observed for the transition vector ${\bm \tau_{10}}$. 
Reactive mutants with strategy $(p^1\leq0.4, q^1; p^2, q^2)~ \forall q^1, p^2, q^2\in[0,1]$ are incapable of invading the Bayesian residents while reactive mutants with strategy defined by $(p^1=1, q^1; p^2, q^2)~ \forall q^1, p^2, q^2\in[0,1]$ could either fully invade or coexist with the Bayesian resident population. 
A key difference from the previous transition vector is that the area of strategy space over which reactive strategies dominate over BTFT increases with increasing $q^1$ even for for $p^1=0.6, 0.8$ as can be seen by comparing Fig.~\ref{fig:ess_q00} and Fig.~\ref{fig:ess_q10}. Hence, the Bayesian strategy can dominate over reactive counterparts that exhibit high reciprocity in the higher benefit state as long as the generosity in that state (characterized by $q^1$) is low.    

When the transition vector is ${\bm \tau_{11}}$, the Bayesian strategy outperform all reactive mutants with strategy $(p^1\leq 0.6, q^1\leq 0.6; p^2, q^2)~ \forall p^2,q^2\in[0,1]$ as can be seen from Fig.~\ref{fig:ess_q11}. When the reciprocity of reactive strategies crosses $p^1=0.6$, the dominance of the BTFT strategy is lost. There exists a sharper threshold ($(p^1\geq 0.8, q^1\geq 0.6; p^2, q^2)~ \forall p^2,q^2\in[0,1]$) for the transition from the dominance of the Bayesian strategy to the dominance of reactive strategies with fewer mixed ESS appearing compared to the previous two transition vectors. Thus, the BTFT strategy fares better against the reactive mutants characterized by lower values of reciprocity ($p^1$) and low generosity ($q^1$) in the beneficial state.

A comparison between the transition vectors ${\bm \tau_{00}}$, ${\bm \tau_{10}}$ and ${\bm \tau_{11}}$, indicate that even in cases where BTFT is an ESS corresponding to regions associated with low reciprocity in the more beneficial game ($p^1 \leq 0.6$), it is not the only ESS. Moreover, differences between the ESS regions for the three transition vectors are most pronounced for $0.6 \leq p^1 \leq 0.8$. This clearly indicates that the transition vector plays an important role in determining when both reactive and Bayesian strategies are evolutionarily stable.

By comparing the area of phase space where BTFT is an ESS, for the three transition vectors, it is evident that the Bayesian strategy performs best against reactive mutants for the transition vector ${\bm \tau_{00}}$. But surprisingly, the Bayesian strategy performs better against reactive mutants in the timeout stochastic game ${\bm \tau_{11}}$ than the timeout game with conditional return ${\bm \tau_{10}}$. 
This can also be realized by comparing the performance of Bayesian against pure strategy reactive mutants. There are 16 such pure reactive strategies, each corresponding to a corner of the four squares---the bottom-left, bottom-right, top-left, and top-right squares---in the ESS phase diagram presented in Figs.~\ref{fig:ess_q00}, \ref{fig:ess_q10}, and \ref{fig:ess_q11}. Out of 16 pure reactive mutants, we notice from Fig.~\ref{fig:ess_q00} that the Bayesian strategy outcompetes 10 reactive mutants in ${\bm \tau_{00}}$ game, while we observe from Fig.~\ref{fig:ess_q10} that the Bayesian strategy outcompetes 9 reactive mutants in the timeout game. In the timeout game with the conditional return, the Bayesian strategy outcompetes 8 pure reactive mutants; see Fig~\ref{fig:ess_q11}.

\subsection{Evolution of cooperation}

\begin{figure}[h!]
	\centering	
	\includegraphics[scale=0.64]{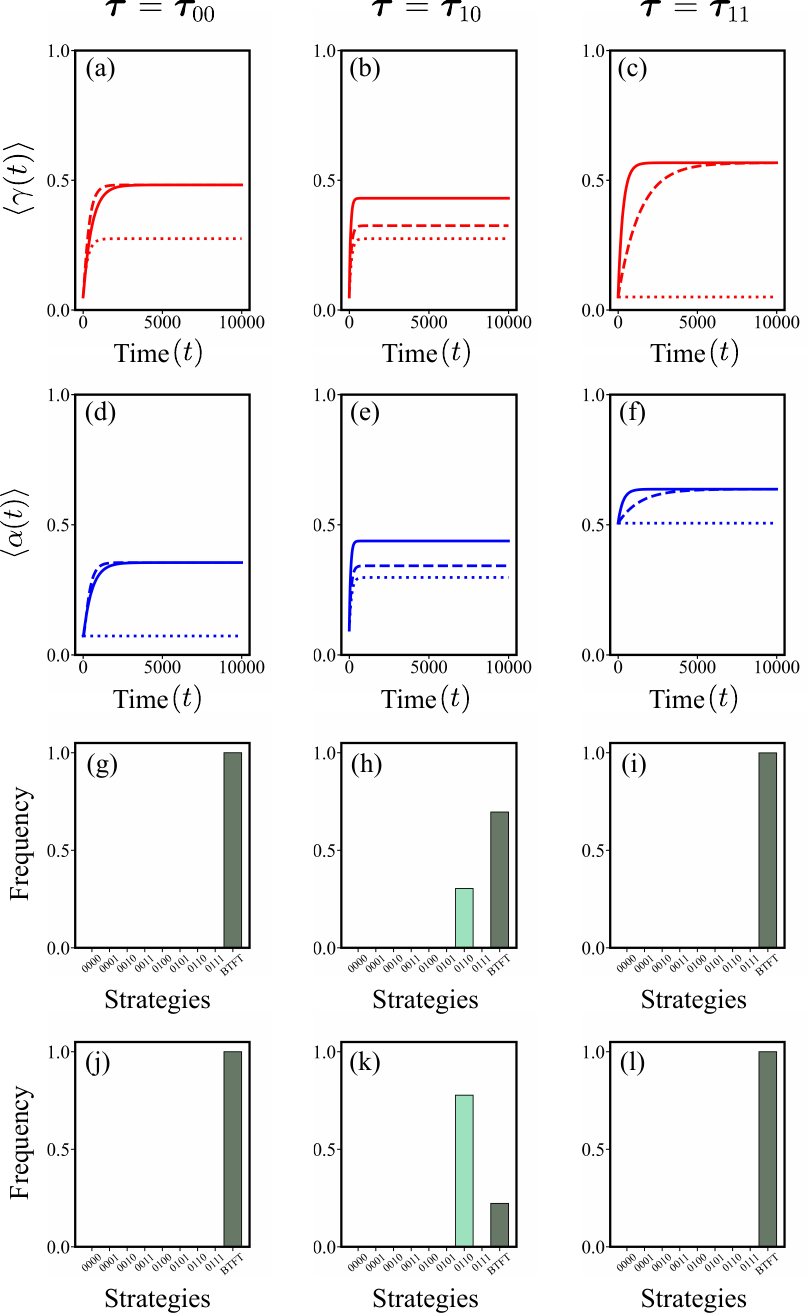}
	\caption{\emph{Evolution of cooperation driven by BTFT:} The average self-cooperation rate and probability of being in the more beneficial game state are shown for two different mutation-selection processes. The first, second, and third columns, respectively, show the results for the transition vectors ${\bm \tau_{00}}$, ${\bm \tau_{10}}$ and ${\bm \tau_{11}}$. The dashed line and the solid line represent the mutation-selection process with the mutation rate of BTFT as  $\frac{1}{9}$ and $\frac{1}{2}$, respectively. The first row (a)-(c) depicts the cooperation level of the population over time, when the population starts from ALLD regardless of the resource's state. The second row (d)-(f) illustrates the frequency of beneficial states over time. The dotted line corresponds to the cooperation level in the absence of BTFT. The third and fourth rows exhibit the frequency of the strategies after $10^4$ generations when the mutation rate of BTFT is  $\frac{1}{2}$ (panels g-i) and $\frac{1}{9}$ (panels j-l), respectively.  Parameters used: $\beta=10$, $N=100$, $\delta=0.9$, $r_1=10$ and $r_2=2$.}
	\label{fig:coop_rate}
\end{figure}

In analyzing the evolutionary dynamics of a population where a resident population can potentially be invaded by a mutant (see Section~\ref{sec:evo_dynamics}), our goal is  understand whether the Bayesian strategy can successfully displace a reactive strategy and then avoid being invaded by other reactive strategies that emerge over time. We also wish to understand how the presence of a Bayesian strategy can affect average cooperation rates and the propensity of the system to remain in the higher benefit state. To address these issues, we consider a set of pure reactive strategies only along with the BTFT strategy. Therefore, there are $2^4+1$ strategies in the setup. We already know from the pioneering work of Axelrod~\cite{Axelrod1981} that TFT is very effective in sustaining cooperation. Since we wish to understand the exclusive role of Bayesian strategy in sustaining cooperation when pitted against more selfish reactive strategies, we exclude the reciprocal pure reactive strategies (TFT and ALLC) in the beneficial state from the strategy set. Our strategy set then consists of 9 strategies including $2^3$ pure reactive strategies and the BTFT strategy.  

We investigate the evolution of the system for two distinct mutation processes. In one case, the mutations are randomly selected from the set of available strategies; therefore, the probability of a given strategy emerging as a mutant is $\frac{1}{9}$. In the other case, the probability of selecting any reactive strategy as the mutant is equal to the probability of selecting the Bayesian strategy as a mutant. Thus, the Bayesian strategy appears as a mutant with probability $\frac{1}{2}$, and each of the $2^3$ reactive strategies appear as a mutant with probability $\frac{1}{16}$. The self-cooperation rate and the probability of getting the beneficial state for these two mutation processes are calculated using Eq.~(\ref{eq:average_cooperation_markov_process}) and Eq.~(\ref{eq:benficial_state_prob_INF}), respectively, upon finding the equilibrium state vector from the Markov process. The results presented in Fig.~\ref{fig:coop_rate} are for the case when the entire population initially follows the ALLD strategy irrespective of the resource state the game starts in.

Fig.~\ref{fig:coop_rate} shows the asymptotic self-cooperation level in the population in the presence (red solid and dashed lines) and absence (red dotted line) of BTFT for the three transition vectors. It is clear that the cooperation levels are much higher in the presence of BTFT with the differences being most pronounced for the timeout game. Similarly, the average propensity to find the game in the more beneficial state is also higher in these two cases. The two different mutation processes represented by the solid and the dashed lines in Fig.~\ref{fig:coop_rate} show similar outcomes for the transition vector are ${\bm \tau_{00}}$ and ${\bm \tau_{11}}$. For the timeout game with conditional return (${\bm \tau_{10}}$), the mutation-selection process, where the Bayesian mutant strategy emerges with a probability $0.5$, leads to higher cooperation levels and higher propensity to be found in the beneficial game state.

We observe that the enhancement of cooperation is solely driven by the BTFT strategy for the transition vector ${\bm \tau_{00}}$ and ${\bm \tau_{11}}$, and it is true for both mutation processes corresponding to $\mu_{S^r S^b}=\frac{1}{2}$  (Fig.~\ref{fig:coop_rate}(g),(i)) and $\mu_{S^r S^b}=\frac{1}{9}$ (Fig.~\ref{fig:coop_rate}(j),(l)). Since the BTFT strategy is an ESS against all available reactive strategies in the set for the transition vector ${\bm \tau_{00}}$ and ${\bm \tau_{11}}$, it is hard to be displaced by any other reactive strategy, once it emerges and gets fixed in the population. However, for the transition vector ${\bm \tau_{10}}$, the reactive strategy (0,1; 1,0) persists and can even dominate BTFT (Fig.~\ref{fig:coop_rate}(k)) depending on the mutation rates. As a result, the enhancement of cooperation is low (Fig.~\ref{fig:coop_rate}(e)). In contrast, the enhancement of cooperation is large when the Bayesian strategy is introduced as a mutant with a higher probability than other potential reactive mutants i.e. $\mu_{S^r S^b}=\frac{1}{2}$, and it is largely driven by the BTFT strategy.


\section{Discussion and Conclusions}

Updating our choices by incorporating new evidences in accordance with Bayes rule is a cognitively demanding task, but one which may have been hardwired and manifest through the potential existence of a Bayesian brain~\cite{Friston2012, Bottemanne2025}. Given the importance of such a framework in both psychology and cognitive neuroscience, it is natural to ask how collective outcomes in the evolution of cooperation are influenced by players employing such Bayesian inferential strategies. Our work establishes a mathematical framework---within the paradigm of evolutionary game theory---for incorporating Bayesian inferential strategies in stochastic games where actions of players can lead to changes in the game (resource) state. 

The success of Bayesian strategies in such games clearly depends on the nature of reactive strategies present in the competing pool as well as on the rule governing the transitions between the two game (resource) states. The BTFT strategy is resilient against the largest set of reactive strategies when the transition rule to the resource rich state is most stringent ($\tau_{00}$), requiring mutual cooperation in the resource poor state. However, the net cooperation rate and the average propensity to be in the resource rich state is higher when the transition rule to the resource rich state is least stringent ($\tau_{11}$) and occurs regardless of the actions of the two players. However, in both these two scenarios, BTFT strategy is most resilient against invasion by other reactive strategies, excluding ALLC and TFT. These exceptions can be attribute to the fact that the Bayesian strategy end up with a lower average payoff when interacting with itself because both Bayesian players occasionally defect with each other. This makes it difficult to satisfy the condition $\pi(S^b, S^b)>\pi(S^r, S^b)$. On the other hand, each TFT can occasionally exploit the BTFT by defecting when the latter cooperates, leading to a lower average payoff for BTFT and making it easier to satisfy the condition $\pi(S^r, S^r)>\pi(S^b, S^r)$ where $S^r=\text{TFT}$. Similarly, since two ALLC players always reap the benefits of mutual cooperation that can outweigh the occasional reduction in payoff when encountering a Bayesian opponent who defects, it is also much easier to satisfy the condition $\pi(S^r, S^r)>\pi(S^b, S^r)$ where $S^r=\text{ALLC}$. It is therefore prudent at this point, to add a note of caution while talking about the efficacy of Bayesian strategies. By being susceptible to invasion by ALLC, the Bayesian strategy opens up the possibility of eventual dominance of ALLD since the latter can invade ALLC. However, such outcomes will be a lot less likely in a mixed population of BTFT and TFT players.

We emphasize that excluding ALLC and TFT from the strategy set does not weaken the conclusions we aim to draw in this paper. We do not ask whether BTFT can outcompete the cooperative strategies TFT and ALLC. Given the paramount importance of the emergence of cooperation in evolutionary game theory, we instead investigate whether the BTFT strategy can survive and promote cooperation in the long term, after it emerges as a single mutant strategy in a population that initially consists of defectors only. In order to investigate it, of course, the most general approach would be to consider BTFT interacting with all pure reactive strategies. In such cases, BTFT does not emerge as the dominant long-term strategy. Again, our goal is to understand the performance of BTFT against always defectors; therefore, we investigate whether this BTFT strategy can survive against the selfish strategy ALLD and enable cooperation in the long term when the cooperative strategies TFT and ALLC are not present in the strategy set. We find that BTFT does indeed take on the role of promoting cooperation when TFT and ALLC are not present in the beneficial state of a stochastic game. 

It is worth mentioning that BTFT could not survive in the long term when transitions between the two resource states are not allowed (see Appendix~\ref{app:cooperation_fixed_payoff_matrix}). In this case, ALLD is the only dominant strategy that survives in the long term (see Appendix G). As a result, the cooperation level in the system is nearly zero. When transitions between the states are switched on, BTFT survives in the long term even if the population initially consists of ALLD. Consequently, BTFT facilitates cooperation, as the overall cooperation level is higher in its presence than in its absence. In summary, the enhancement of cooperation arises from the interplay between the stochastic game transitions and the BTFT strategy even though neither of them are solely capable of fostering cooperation.

Our analysis of Bayesian inferential strategies in the context of stochastic games differs in subtle ways from our previous work~\cite{PatraNJP2024} on the efficacy of Bayesian strategies in a single game. In the current work, the player employing a Bayesian strategy can update her beliefs about the reactive opponent's strategy in {\it both} games on the basis of evidence collected from the opponent's actions over time.  A recent analysis~\cite{Kleshnina2023} has suggested that incomplete information can impact cooperation levels in stochastic games. Hence it would be interesting to see how allowing for complete information, about both actions and resource states in which those actions were taken, in Bayesian belief updating process can affect the outcome. Another natural extension of our work involves the outcome of competition between a Bayesian strategy and the more cognitively demanding memory-one strategies in the context of stochastic games. However, incorporating memory-one strategies into the Bayesian framework significantly increases computational complexity: the Bayesian updating process would require handling 8-dimensional prior and posterior distributions. One envisages that if such simulations are performed,  one would expect to witness the defining role of the Win-Stay-Lose-Shift (WSLS) strategy in conjunction with the Bayesian strategies in fostering cooperation. 

Even though several experiments~\cite{Mazalov1996,Luttbeg1996,Welton2003, JValone2006, Biernaskie2009} suggest that both humans and other animals take decisions by using new evidence in a manner that is consistent with Bayes rule, systematic deviations from Bayesian inference have also been observed \cite{Phillips-Hays-Edwards1966,Phillips-Edwards1966} in many experiments involving humans. Those deviations can be attributed to specific cognitive biases that leads to under-weighing new evidence and/or under-weighing the prior (also called base-rate neglect \cite{achtziger2014}. It is important to understand whether accounting for such cognitive biases can lead to significantly different outcomes in social dilemmas. Bayesian strategies, along with modified counterparts that incorporate human cognitive biases, can be crucial for the emergence and resilience of collective altruism since they incorporate learning in realistic ways that go beyond simple heuristics that are the hallmark of reactive strategies. We, therefore, hope that our work will motivate further exploration of Bayesian inferential strategies in the context of evolution of cooperation.

\begin{acknowledgments}
AP thanks CSIR (India) for the financial support in the form of Senior Research Fellowship. Authors thank Upayan Roy for helpful discussions.
\end{acknowledgments}

\appendix

\section{Likelihood}
\label{app:lhf}

Let us consider a two-state-two-action-two-player stochastic game where the focal player uses the reactive strategy ($p^1, q^1; p^2, q^2$), and the opponent has the strategy ($\tilde{p}^1, \tilde{q}^1; \tilde{p}^2, \tilde{q}^2$). Therefore, the Markov chain that describes the repeated interaction in stochastic game corresponds to total eight states: ($s^1, C, C$), ($s^1, C, D$), ($s^1, D, C$), ($s^1, D, D$), ($s^2, C, C$), ($s^2, C, D$), ($s^2, D, C$), and ($s^2, D, D$). In each 3-tuple, the first parameter denotes the environmental states; whereas the second and the third elements correspond to the action of the focal and opponent, respectively. In a compact notation, one can write these states as $\omega=(s^i, a, \tilde{a})$ where $s^i\in\{s^1, s^2\}$ denotes the environmental states, $a$ and $\tilde{a}$ are the actions of focal and opponent players, respectively. Thus, the Markov chain can be written as follows,
\begin{eqnarray}
\sigma^i_{ CC, n}&=&\smashoperator[r]{\sum_{\makecell{a, \tilde{a}\in\{C, D\}\\ j=1,2}}}P(s^i,C,C|s^j,a,\tilde{a})\sigma^j_{a\tilde{a},n-1},\nonumber\\
\sigma^i_{ CD, n}&=&\smashoperator[r]{\sum_{\makecell{a, \tilde{a}\in\{C, D\}\\ j=1,2}}}P(s^i,C,D|s^j,a,\tilde{a})\sigma^j_{a\tilde{a},n-1},\nonumber\\
\sigma^i_{ DC, n}&=&\smashoperator[r]{\sum_{\makecell{a, \tilde{a}\in\{C, D\}\\ j=1,2}}}P(s^i,D,C|s^j,a,\tilde{a})\sigma^j_{a\tilde{a},n-1},\nonumber\\
\sigma^i_{ DD, n}&=&\smashoperator[r]{\sum_{\makecell{a, \tilde{a}\in\{C, D\}\\ j=1,2}}}P(s^i,D,D|s^j,a,\tilde{a})\sigma^j_{a\tilde{a},n-1}.\nonumber
\end{eqnarray} 
Note that the probability of cooperation of a player in state $i$ in $n$-th round, $P_n(s^i,C)$ can be found from the above Markov chain by using the following relation: $P_n(s^i,C)=\sigma^i_{ CC, n}+\sigma^i_{ CD, n}$. Now, we illustrate the calculation for the probability of cooperation of the focal player in state $s^1$,  and the rest of them follow similar steps. 

The probability, $P(s^1,C,C|s^j,C,C)$, is found by multiplying the transition probability from the state $s^j$ to $s^1$ with the probability of the cooperation of focal and opponent players given the past actions of the players $CC$ in state $j$. Therefore, it is $\tau^j_{CC}p^i\tilde{p}^i$. Similarly, other conditional probabilities can be found by employing the players' strategies and the transition rule between the states. In order to find the expression for the unconditional cooperation probability of the focal player, we find the probability of state ($s^1, C,C$), which is,
\begin{eqnarray}
	\sigma^1_{ CC, n}&=& \tau^1_{CC}p^1\tilde{p}^1\sigma^1_{ CC, n-1}+\tau^1_{CD}q^1\tilde{p}^1\sigma^1_{ CD, n-1}\nonumber\\&&+\tau^1_{DC}p^1\tilde{q}^1\sigma^1_{ DC, n-1}+\tau^1_{DD}q^1\tilde{q}^1\sigma^1_{ DD, n-1}\nonumber\\&&+\tau^2_{CC}p^1\tilde{p}^1\sigma^2_{ CC, n-1}+\tau^2_{CD}q^1\tilde{p}^1\sigma^2_{ CD, n-1}\nonumber\\&&+\tau^2_{DC}p^1\tilde{q}^1\sigma^2_{ DC, n-1}+\tau^2_{DD}q^1\tilde{q}^1\sigma^2_{ DD, n-1}\nonumber.\\\label{eq:a1}
\end{eqnarray}
In a similar manner, the probability of state ($s^1, C, D$) is
\begin{eqnarray}
	\sigma^1_{ CD, n}&=&\tau^1_{CC}p^1(1-\tilde{p}^1)\sigma^1_{ CC, n-1}+\tau^1_{CD}q^1(1-\tilde{p}^1)\sigma^1_{ CD, n-1}\nonumber\\&+&\tau^1_{DC}p^1(1-\tilde{q}^1)\sigma^1_{ DC, n-1}+\tau^1_{DD}q^1(1-\tilde{q}^1)\sigma^1_{ DD, n-1}\nonumber\\&+&\tau^2_{CC}p^1(1-\tilde{p}^1)\sigma^2_{ CC, n-1}+\tau^2_{CD}q^1(1-\tilde{p}^1)\sigma^2_{ CD, n-1}\nonumber\\&+&\tau^2_{DC}p^1(1-\tilde{q}^1)\sigma^2_{ DC, n-1}+\tau^2_{DD}q^1(1-\tilde{q}^1)\sigma^2_{ DD, n-1}. \nonumber\\
	\label{eq:a2}
\end{eqnarray}
Therefore, the cooperation probability of unprimed player in state $s^1$ using Eq.~(\ref{eq:a1}) and Eq.~(\ref{eq:a2}), is
\begin{eqnarray}
	P_n(s^1,C)&=&\sigma^1_{ CC, n}+\sigma^1_{ CD, n}\nonumber,\\
	&=&\tau^1_{CC}p^1\sigma^1_{ CC, n-1}+\tau^1_{CD}q^1\sigma^1_{ CD, n-1}\nonumber\\&&+\tau^1_{DC}p^1\sigma^1_{ DC, n-1}+\tau^1_{DD}q^1\sigma^1_{ DD, n-1}\nonumber\\&&+\tau^2_{CC}p^1\sigma^2_{ CC, n-1}+\tau^2_{CD}q^1\sigma^2_{ CD, n-1}\nonumber\\&&+\tau^2_{DC}p^1\sigma^2_{ DC, n-1}+\tau^2_{DD}q^1\sigma^2_{ DD, n-1}. \nonumber\\\label{eqn:unconditional_prob_focal_unorganised}
\end{eqnarray}
On rearranging Eq.~(\ref{eqn:unconditional_prob_focal_unorganised}),  we arrive at the expression for likelihood in Eq.~\ref{eq:s1C}. Similar calculations can be done to obtain the likelihoods $P_n(s^1,D)$, $P_n(s^2,C)$ and $P_n(s^2,D)$ for other possible evidences $(s^1, D)$, $(s^2, C)$ and $(s^2, D)$.

\section{Standard corner strategies}
\label{app: true_corner}
\begin{figure*}
	\centering	
	\includegraphics[scale=0.85]{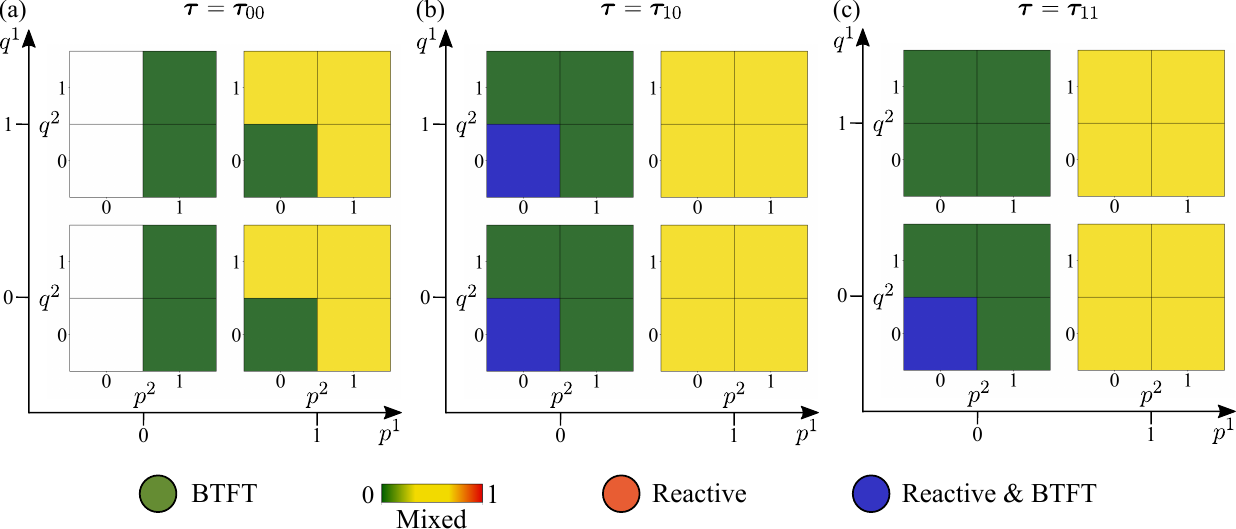}
	\caption{ESS phase diagram for BTFT vs. standard corner reactive strategies in infinite population for three transition vectors: ${\bm \tau_{00}}$, ${\bm \tau_{10}}$ and ${\bm \tau_{11}}$. White regions represent corner strategies for which the Bayesian updating process fails, and therefore, no definitive outcomes for BTFT can be determined against these strategies.}
	\label{fig:true_corner}
\end{figure*}
It is usually followed in the literature that corner reactive strategies have a specific action in the initial round. For example, the ALLD strategy begins with defection, while ALLC and TFT start with cooperation. When this criterion holds, we refer to these as standard corner strategies. However, the initial state vector $\bm{\sigma}_1$ used in the main text does not adhere to this criterion. Instead, we assume that players are indifferent about which action to take in the first round and therefore cooperate with probability 0.5, following the principle of insufficient reason. If we construct the initial state vector $\bm{\sigma}_1$ based on the standard definition of corner reactive strategies, and the game starts from the beneficial resource with probability 0.5, the Bayesian updating process fails (the likelihood becomes zero for some evidences) for the transition vector $\bm \tau_{00}$ for the four corner strategies, viz.,  $0,0;0,0$, $(0,0; 0,1)$, $(0,1; 0,0)$, and $(0,1; 0,1)$. These strategies are represented in white in Fig.~\ref{fig:true_corner}. Note that we assume Anti-TFT ($p^i=0$ and $q^i=1$) defects in the initial round, and that Bayesian players always apply the principle of insufficient reason, cooperating with probability 0.5 initially. The updating process functions correctly (the likelihoods don't vanish for some evidences) for each standard corner reactive strategy in the other two transition vectors, $\bm \tau_{00}$ and $\bm \tau_{11}$, and the corresponding outcomes are showcased in Fig.~\ref{fig:true_corner}.

In the transition vector $\tau_{00}$, out of remaining 12 corner strategies (recall: for 4 corner strategies Bayesian updating fails), the BTFT is only ESS against six corner strategies. It means that the BTFT mutant fully invades the population carrying any one of these 6 strategies. It is shown by the green color in Fig~\ref{fig:true_corner}(a). However, the BTFT mutant can not fully invade the population, adopting the other 6 corner strategies and coexists with them---it is represented by the yellow color. For the transition vector $\bm \tau_{10}$ and $\bm \tau_{11}$, the (ALLD; ALLD) mutant could not out-compete the BTFT residents and vice versa. Notice that this result is similar to the outcomes in the main text. In addition, the BTFT strategy is ESS against 8 reactive corner strategies among 16 corner strategies for these two transition vectors. The Bayesian strategy coexists with the other 8 corner strategies. The difference in outcomes between the transition vectors $\bm \tau_{10}$ and $\bm \tau_{11}$ is that (Anti-TFT; ALLD) is ESS for $\bm \tau_{10}$ while it is not for $\bm \tau_{11}$. The crucial point here is to make that the mutants with standard corner reactive strategies never fully invade the BTFT resident population, while the pure reactive mutants in the main text can fully invade the BTFT residents; for example, compare the outcomes for the corner (ALLC; ALLC) in Fig.~\ref{fig:ess_q00}, Fig.~\ref{fig:ess_q10}, Fig.~\ref{fig:ess_q11} and Fig.~\ref{fig:true_corner}. Thus, the evolutionary performance of BTFT is even better against the standard corner reactive strategies than against the corner reactive strategies used in the main text.

\section{Finding payoff matrix numerically}
\label{app:finding_Payoff matrix_numerically}

$\pi(S^r,S^r)$ can be easily determined following the description given in Sec.~\ref{sec:sec_ES_BTFT}. However, the other elements of the payoff matrix ${\sf \Pi}$ require more careful computations. Interactions involving a Bayesian player requires specifying how the posterior distribution is numerically calculated.
	
	To simulate the repeated interaction of Bayesian player, we take an uniform prior distribution at $n=1$ over the sample space $p^1$-$q^1$-$p^2$-$q^2$. {In our simulations, $p^i$s and $q^i$s are independently sampled in steps of $0.2$ from closed interval $[0,1]$. This amounts to discretizing the sample space into $6^4$ grid points and make an array with $6^4$ elements where each element is assigned equal probability.} 
    The posterior distribution is determined at the end of each round using Eq.~(\ref{eq:Bayes}). Naturally, the posterior distribution is also  a distribution over $6^4$ grid points and is used as the prior in the subsequent round. The maxima of the posterior at the end of a round is used as the BTFT player's strategy in the subsequent round and therefore informs her actions in the subsequent round. Although theoretically a repeated game goes on \emph{ad infinitum}, for the sake of numerical calculations, we ran each trail of repeated game for $n_f=100$ rounds while choosing the discount factor $\delta=0.9$. The resultant total accumulated payoffs are then computed using Eq.~(\ref{eq:Average_payoff_calculation}) and averaged over $10^5$ independent trials.
	
In order to decide on appropriate number of digits to which the payoffs' values should be truncated, we calculate the standard deviation of any (averaged) payoff element (\emph{sample mean}) of ${\sf \Pi}$; the central limit theorem dictates that it be of the order $sd/\sqrt{10^{5}}$, where $sd$ is the standard deviation of the independent identical random accumulated payoffs (\emph{samples}) in different trails.  We numerically find $sd$ and conclude that $sd/\sqrt{10^{5}}\sim 10^{-1}$ (also verified by direct numerical estimation of standard deviation of the sample mean). Consequently, we decide to round off the average payoffs up to the first place of decimal. 
 
\section{Imhof--Nowak--Fudenberg process}
\label{app:INF_process}
To study the long-term evolutionary dynamics in a well-mixed finite population, we employ the Imhof–Nowak–Fudenberg process. In this framework, a rare mutant arises in an monomorphic population at each time step and either goes extinct or successfully invades. In both cases, the population returns to a monomorphic state---either adopting population-wide mutant strategy (if invasion succeeds) or sticking with the resident strategy. Thus, the population undergoes successive transitions between monomorphic states on a long time scale. However, before the mutant either establishes itself or dies out, the population consists of two types—resident and mutant—representing a transient state.

In the transient state, the selection between the resident and mutant in the population is governed by the imitation process, or it can also be viewed as a Moran process. A focal player in the population can randomly meet with either residents or mutants and receives an expected payoff $\pi$; then she compares the expected payoff with a randomly chosen individual's expected payoff, $\tilde{\pi}$. Next the focal player changes her strategy to the randomly chosen individual's strategy with a probability $(1+ \exp[-\beta(\tilde{\pi}-\pi)])^{-1}$, where $\beta$ represents the selection strength. The strength of selection is weak when $\beta\rightarrow 0$ and strong when $\beta\rightarrow\infty$. At the end of the selection process, the population is in a homogeneous state corresponding to either resident or mutant. 

The process can be studied analytically using a discrete-state discrete-time Markov chain~\cite{Fudenberg2006_JET}. The states of the Markov chain correspond to the strategies available in the system. The introduction of a rare mutant can lead to transition between distinct states of the Markov chain. 
The fixation probability of mutant strategy ${S}_{\text{mut}}$ within the residents of strategy ${S}_{\text{res}}$ is denoted as $\rho({S}_{\text{res}},{S}_{\text{mut}})$. With probability $\rho({S}_{\text{res}},{S}_{\text{mut}})$, the mutant fixates and becomes a new resident in the population. Otherwise, it goes extinct with probability $1-\rho({S}_{\text{res}},{S}_{\text{mut}})$, and the resident population remains unchanged. 

To find an expression of fixation probability $\rho({S}_{\text{res}},{S}_{\text{mut}})$, we assume that there are $k$ mutants of strategy ${S}_{\text{mut}}$ and $(N-k)$ residents of strategy ${S}_{\text{res}}$ in the population. Then the expected payoffs of the resident and mutant are respectively given by 
\begin{equation}
	\pi_{\text{res}}(k)=\frac{N-k-1}{N-1}\pi({S}_{\text{res}},{S}_{\text{res}})+\frac{k}{N-1}\pi({S}_{\text{res}},{S}_{\text{mut}}),
\end{equation} 
\begin{equation}
	\pi_{\text{mut}}(k)=\frac{N-k}{N-1}\pi({S}_{\text{mut}},{S}_{\text{res}})+\frac{k-1}{N-1}\pi({S}_{\text{mut}},{S}_{\text{mut}}).
\end{equation}%
where ${S}_{\text{res}}$ and  ${S}_{\text{mut}}$  can be any strategy from the fixed set, $\mathcal{S}$, containing the BTFT strategy $S^b$ and countably many reactive strategies as desired; $\pi({S}_{\text{res}},{S}_{\text{res}})$, $\pi({S}_{\text{res}},{S}_{\text{mut}})$, $\pi({S}_{\text{mut}},{S}_{\text{res}})$, and $\pi({S}_{\text{mut}},{S}_{\text{mut}})$ can be obtained from using Eq.~(\ref{eq:Average_payoff_calculation}) or  Eq.~(\ref{eq:wighted_sum_payoff}), whichever is appropriate. It is a textbook knowledge that the fixation probability of a mutant with strategy ${S}_{\text{mut}}$ within the resident population of strategy ${S}_{\text{res}}$ must be
\begin{equation}
	\rho({S}_{\text{res}},{S}_{\text{mut}})=\frac{1}{1+\sum_{i=1}^{N-1}\prod_{k=1}^{i} e^{-\beta[\pi_{\text{mut}}(k)-\pi_{\text{res}}(k)]}}.
\end{equation}

Finally, we construct the transition matrix, {\sf R},where the transition probabilities can be found using the fixation probabilities and mutation rates. The transition probability, $r({S}, \tilde{S})$, from a current resident strategy ${S}$ to the next resident strategy $\tilde{ S}$ can be expressed as
\begin{equation}
	r({S}, \tilde{S}) =
	\begin{cases}
		\mu_{{S}\tilde{S}}\rho({S}, \tilde{S})~~~~~~~~~~~~~~~~~~\text{if}~{S}\neq\tilde{S}, & \\
		1- \sum_{{\substack{\tilde{ S}\in\mathcal{S}\\\tilde{S}\neq {S}}}}\mu_{{S}\tilde{S}}\rho({S}, \tilde{S})~~~~ \text{if}~{S}={\bm \tilde{ S}}. & 
	\end{cases}   
\end{equation}%
{Here, $\mu_{{S}\tilde{S}}$ is the rate at which the strategy $\tilde{ S}$ appears in the population where the resident strategy is ${S}$.} For example, $\mu_{{S}\tilde{S}}=\frac{1}{|\mathcal{S}|}$ when the mutants are uniformly randomly selected from the set of available strategies $\mathcal{S}$ and $\mathcal{|S|}$ denotes the number of strategies in the set. The state vector $\bm x(t)$ made up of $x_S(t)$s as components at time step $t$ can be easily found from the Markov process: $\bm x(t)=\bm x(0){\sf R}^t.$
\section{Cooperation rate of reactive strategy}
\label{app:cooperation_rate_reactive_strategy}
When two reactive players with strategies ${S^r}$ and $\tilde{S}^r$ interact repeatedly, the game state can change depending on the actions of both the players. We use the framework present in literature~\cite{Hilbe2018Nature} to obtain the cooperation rate of a focal reactive player. It involves calculating the equilibrium probability of finding the game between the two reactive players in each of the 8 possible states of a Markov chain denoted by $\omega=(s^i, a, \tilde{a})$. 

The entire process of interaction can be described in terms of transitions between the different possible states of the Markov chain. The transition probability between these states is determined by the transition rule, ${\bm \tau}$, between the resource states and the transition rule the strategies, ${S^r}$ and $\tilde{S}^r$, of the players. Thus, the transition probability from the states $\omega=(s^i, a, \tilde{a})$ to the state $\omega'=(s^{i'}, a', \tilde{a}')$ may be conveniently represented as 
\begin{equation}
	m_{\omega, \omega'}=z \cdot y\cdot\tilde{y}.
\end{equation}
The first factor $z$ describes the probability of transition to the current resource state $s^{i'}$ from the former resource state $s^i$, and it is governed by the transition vector ${\bm \tau}$. The probability $z$ can be found as follows
\begin{equation}
	z =
	\begin{cases}
		\tau^{i}_{a\tilde{a}}~~~~~~~~~\text{if}~s^{i'}=s^1, & \\
		1-\tau^{i}_{a\tilde{a}}~~~~ \text{if}~s^{i'}=s^2. & 
	\end{cases}   
\end{equation}
The second and third factors $y$ and $y'$ are the conditional probabilities of playing the action $a'$ and $\tilde{a}'$ of player-1 and player-2, respectively, given the last action pair $a\tilde{a}$  and the current state $s^{i'}$. These factors are decided by the strategies ${S^r}$ and $\tilde{S}^r$ of player-1 and player-2, respectively. The factor $y$ can be determined as follows:

\begin{equation}
	y =
	\begin{cases}
		p^{i'(r)}~~~~~~~~~\text{if}~a'=C~\text{and}~\tilde{a}=C, & \\
		1-p^{i'(r)}~~~~ \text{if}~a'=D~\text{and}~\tilde{a}=C, & \\
		q^{i'(r)}~~~~~~~~~\text{if}~a'=C~\text{and}~\tilde{a}=D,&\\
		1-q^{i'(r)}~~~~ \text{if}~a'=D~\text{and}~\tilde{a}=D.
	\end{cases}   
\end{equation}
Similarly, the factor $\tilde{y}$ can be calculated as follows:
\begin{equation}
	\tilde{y} =
	\begin{cases}
		\tilde{p}^{i'(r)}~~~~~~~~~\text{if}~\tilde{a}'=C~\text{and}~a=C, & \\
		1-\tilde{p}^{i'(r)}~~~~ \text{if}~\tilde{a}'=D~\text{and}~a=C, & \\
		\tilde{q}^{i'(r)}~~~~~~~~~ \text{if}~\tilde{a}'=C~\text{and}~a=D,&\\
		1-\tilde{q}^{i'(r)}~~~~  \text{if}~\tilde{a}'=D~\text{and}~a=D.
	\end{cases}   
\end{equation}
Thus, following the above procedure, we obtain all elements $m_{\omega, \omega'}$ of the transition matrix ${\sf M}({S^r}, \tilde{ S}^r)$ of the Markov chain.

For stochastic games between two reactive strategies with discounting, the cooperation rate between two players can be calculated analytically using the transition matrix and the initial state vector of the Markov chain. We assume that ${\bm \sigma}_1$ is the initial probability of finding the system in each of the 8 possible states of the Markov chain such that the component $\sigma^i_{a\tilde{a}, 1}$ represents the probability that player-1 and player-2, respectively, take actions $a$ and $\tilde{a}$ in the initial resource state $s^{i}$. Then, the weighted average state vector can be derived as follows,
\begin{equation}
	\bar{\bm \sigma}=\frac{\sum_{n=1}^{\infty} {\bm \sigma}_1(\delta{\sf M})^{n-1}}{\sum_{n=1}^{\infty}\delta^{n-1}}=(1-\delta){\bm \sigma}_1({\sf I}-\delta{\sf M})^{-1}.
	\label{eq:wighted_averaged_state_vector}
\end{equation}
where $\sf I$ is the identity matrix of size $8\times8$. The elements $\bar{\sigma}^{i}_{a\tilde{a}}$ can be interpreted as the probabilities of observing the states $\omega=(s^i, a, \tilde{a})$ at the end of the game with effective game length $\frac{1}{1-\delta}$.

Now, we can calculate the rate of cooperation for the pair of strategies ${S^r}$ and $\tilde{ S}^r$ from the weighted average state vector. The cooperation rate is following 
\begin{equation}
	\gamma({S^r}, \tilde{S}^r)=\sum_{i\in \{1,2\}}\left[\bar{\sigma}^i_{CC}+\frac{\bar{\sigma}^i_{CD}+\bar{\sigma}^i_{DC}}{2}\right].
	\label{eq:cooperation_rate}
\end{equation}%
The quantities $\gamma({S}, \tilde{ S}^r)$ and $\gamma(\tilde{S}^r, {S})$ are always equal, and the self-cooperation rate of the strategy ${S}$ is given by $\gamma({S}, {S})$. The probability of being in the beneficial state (corresponding to $i=1$) can be found from the following equation:
\begin{equation}
	\alpha({S^r}, \tilde{S}^r)=\sum_{a,\tilde{a}\in\{C,D\}} \bar{\sigma}^1_{a\tilde{a}}.
	\label{eq:beneficial_state_prob}
\end{equation}%
Finally, we point out that the payoff, which a focal reactive player with ${S^r}$ strategy  gets when playing against her opponent, another reactive player with strategy $\tilde{S}^r$, can also be found using $\bar{\bm \sigma}$ through the following formula:
\begin{equation}
	\pi({S^r}, \tilde{S}^r)=\sum_{i\in \{1,2\}}\left[r_i\left(\sum_{a\in \{C,D\}}\bar{\sigma}^i_{aC}\right)-\left(\sum_{\tilde{a}\in \{C,D\}}\bar{\sigma}^i_{C\tilde{a}}\right)\right].
	\label{eq:wighted_sum_payoff}
\end{equation}%

\section{Self-cooperation rate of BTFT}
\label{app:cooperation_rate_BTFT}
As the BTFT player samples a large number of reactive strategies over rounds, finding the weighted average state vector from Eq.~(\ref{eq:wighted_averaged_state_vector})---which assumes two fixed reactive strategies playing against each other---is not feasible. Thus, we find the self-cooperation rate of the BTFT player using simulations, as described below.

Since an element of the weighted averaged state vector $\bar{\bm \sigma}$ can be interpreted as the frequency of observing a state $\omega=(s^i, a, \tilde{a})$ within the effective game length $\frac{1}{1-\delta}$, we consider the repeated game between the two {Bayesian} players up to the effective game length $n_f=\frac{1}{1-\delta}$. Subsequently, we record how often ($\#\omega$) each state $\omega=(s^i, a, \tilde{a})$ appears over $n_f$ rounds and divide those numbers by $n_f$ to get a weighted state vector: $i.e.$, $\sigma^{i}_{a\tilde{a}}=\frac{\#\omega}{n_f}$. {Note that there are 8 distinct values of $\sigma^{i}_{a\tilde{a}}$ since $i={1,2}$ and $a,\tilde{a} \in {C,D}$. Each of the eight} weighted state vectors ($\sigma^{i}_{a\tilde{a}}$) is random when repeated over separate trials; thus, we compute the averaged weighted state vector by averaging over $N_T=10^5$ trials to get 
	\begin{equation}
		\bar{\sigma}^{i}_{a\tilde{a}}=\frac{\sum\limits_{i=1}^{i=N_T}{\sigma^{i}_{a\tilde{a}}}}{N_T}.
	\end{equation}
The self-cooperation rate {$\gamma(S^b, S^b)$} of the BTFT player can then be calculated from the relation $\gamma(S^b, S^b)=\sum\limits_{i\in\{1,2\}}\left[\bar{\sigma}^i_{CC}+\frac{\bar{\sigma}^i_{CD}+\bar{\sigma}^i_{DC}}{2}\right]$. The probability of being in the beneficial state when two Bayesian players interact can also be found from the weighted average state vector using the relation: $\alpha(S^b, S^b)=\sum\limits_{a,\tilde{a}\in\{C,D\}} \bar{\sigma}^1_{a\tilde{a}}$. It is worth mentioning that the realized cooperation level and the probability of being in the beneficial state depend on the choice of transition vector. The realized cooperation levels are 0.482, 0.497, and 0.567 for the transition vectors ${\bm \tau_{00}}$, ${\bm \tau_{10}}$, and ${\bm \tau_{11}}$, respectively (rounded to three decimal places). The corresponding probabilities of being in the beneficial state are 0.355, 0.499, and 0.637 for ${\bm \tau_{00}}$, ${\bm \tau_{10}}$, and ${\bm \tau_{11}}$, respectively (rounded to three decimal places).


\section{Cooperation rate in a fixed state}
\label{app:cooperation_fixed_payoff_matrix}
\begin{figure}
	\centering	
	\includegraphics[scale=0.43]{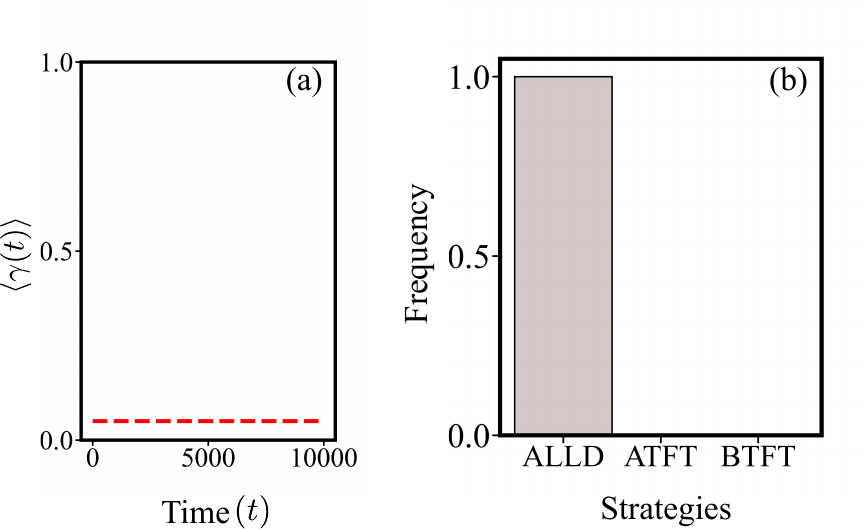}
	\caption{The average self-cooperation rate is shown when transitions between the states are now allowed. Subplot (b) represents the frequency of strategies after $10^4$ generation. The game was fixed in beneficial state to generate this plot. Parameter values: $\beta=10$, $N=100$, $\delta=0.9$, $r_1=10$. }
	\label{fig:coop_rate_q1100}
\end{figure}
Suppose transitions between the resource states are turned off, meaning that the underlying state, and hence the payoff matrix, remains fixed over time. We want to know the asymptotic cooperation level of the system in such a limiting case. Mathematically, in the framework of the stochastic game used in this paper, transition vector $\bm{\tau} = (1,1,1;0,0,0)$ ensures that no transitions occur between the two states.  As in the main text, we exclude the reciprocal pure reactive strategies (TFT and ALLC).

To determine the asymptotic cooperation level, we fix the game in state $s^1$ without any loss of generality and present the corresponding cooperation level and the probability of being in the beneficial state in Fig.~\ref{fig:coop_rate_q1100}. The asymptotic cooperation level is observed to be nearly zero (see Fig.~\ref{fig:coop_rate_q1100}(a)). This occurs because, in the long-term evolutionary dynamics, only the ALLD strategy survives in the beneficial state (see Fig.~\ref{fig:coop_rate_q1100}(b)). Thus, comparing Fig.~\ref{fig:coop_rate_q1100} with Fig.~\ref{fig:coop_rate}, we find that the propensity for cooperation is higher when transitions between resource states are allowed than when they are not; this higher level of cooperation arises due to the persistence of BTFT in the long-term evolutionary dynamics.

\bibliography{patra_etal_references}    
\end{document}